\documentclass[apjl,twocolumn]{emulateapj_mod}
\usepackage{epsfig,apjfonts,mathptmx}

\def\gtsima{$\; \buildrel > \over \sim \;$}
\def\ltsima{$\; \buildrel < \over \sim \;$}
\def\prosima{$\; \buildrel \propto \over \sim \;$}
\def\gsim{\lower.5ex\hbox{\gtsima}}
\def\lsim{\lower.5ex\hbox{\ltsima}}
\def\simgt{\lower.5ex\hbox{\gtsima}}
\def\simlt{\lower.5ex\hbox{\ltsima}}
\def\simpr{\lower.5ex\hbox{\prosima}}

\def\h1{$h^{-1}$}
\def\eeq{\end{equation}}
\def\beq{\begin{equation}}
\def\24mu{24\,$\mu{\rm m}$}
\def\70mu{70\,$\mu{\rm m}$}
\def\8mu{8\,$\mu{\rm m}$}

\shorttitle{Molecular gas in $z=1.5$ disk galaxies}

\shortauthors{E. Daddi et al.}

\begin{document}

\title{
Very High Gas Fractions and Extended Gas Reservoirs in 
$\lowercase{z}=1.5$ disk galaxies
}

   \author{
	E. Daddi\altaffilmark{1},
          F. Bournaud
          \altaffilmark{1},
          F. Walter
          \altaffilmark{2},
          H. Dannerbauer
          \altaffilmark{1},
          C.L. Carilli
          \altaffilmark{3},
          M. Dickinson
          \altaffilmark{4},
          D. Elbaz
          \altaffilmark{1},
          G.E. Morrison
          \altaffilmark{5,6},
          D. Riechers
          \altaffilmark{7},
          M. Onodera
          \altaffilmark{1},
          F. Salmi
          \altaffilmark{1},
          M. Krips
          \altaffilmark{8},
          D. Stern
          \altaffilmark{9}
           }

\altaffiltext{1}{Laboratoire AIM, CEA/DSM - CNRS - Universit\'e Paris Diderot,
       DAPNIA/Service d'Astrophysique, CEA Saclay, Orme des Merisiers,  91191 Gif-sur-Yvette Cedex, France
    [e-mail: {\em edaddi@cea.fr}]}
\altaffiltext{2}{Max-Planck-Institut f\"ur Astronomie, K\"onigstuhl 17, D-69117 Heidelberg, Germany}
\altaffiltext{3}{National Radio Astronomy Observatory, P.O. Box 0, Socorro, NM 87801}
\altaffiltext{4}{National Optical Astronomy Observatory,  950 N. Cherry Ave., Tucson, AZ, 85719}
\altaffiltext{5}{Institute for Astronomy, University of Hawaii, Honolulu, HI 96822}
\altaffiltext{6}{Canada-France-Hawaii Telescope, Kamuela, HI 96743}
\altaffiltext{7}{Caltech, Pasadena, CA 91109}
\altaffiltext{8}{Institut de Radio Astronomie Millim\'etrique (IRAM), St. Martin d'H\`eres, France}
\altaffiltext{9}{Jet Propulsion Laboratory, California Institute of Technology, Pasadena, CA 91109}

\begin{abstract}

We present evidence for very high gas fractions and extended molecular gas reservoirs in normal, near-infrared
selected (BzK) galaxies at z$\sim$1.5. Our results are based on multi-configuration CO[2-1] observations obtained
at the IRAM Plateau de Bure Interferometer. All six star forming
galaxies observed were detected at high significance. High
spatial resolution observations resolve the CO emission in four of them, implying sizes of the gas reservoirs  of
order of 6--11~kpc and suggesting the presence of ordered rotation. The galaxies have UV morphologies consistent with
clumpy, unstable disks, and UV sizes that are consistent with those measured in CO. The star formation efficiencies are homogeneously low within the sample and similar to those of local spirals -- 
the resulting gas depletion times are $\sim0.5$~Gyr, much higher than what is seen in high-z submm galaxies and quasars. 
The CO luminosities can be predicted to within 0.15~dex from the observed star formation rates and stellar masses, 
implying a tight correlation of the gas mass with these quantities.
We use new dynamical models of clumpy disk galaxies to derive dynamical masses for our sample. These models are able
to reproduce the peculiar spectral line shapes of the CO emission. After accounting for the stellar and dark matter 
masses we derive molecular gas reservoirs with masses of 0.4--1.2$\times10^{11}M_\odot$.
The implied conversion (CO luminosity-to-gas mass) factor is very high: $\alpha_{\rm CO}=3.6\pm0.8$, consistent with
a Galactic conversion factor but four times higher than that of local ultra-luminous IR galaxies that is typically used
for high-redshift objects. The gas mass in these galaxies is comparable to or larger than the stellar mass, and the
gas accounts for an impressive 50--65\% of the baryons within the galaxies' half light radii. We are thus
witnessing truly gas-dominated galaxies at $z\sim1.5$, a finding that explains the high specific SFRs observed for
$z>1$ galaxies. The BzK galaxies can be viewed as scaled-up versions of local disk galaxies, with low efficiency
star formation taking place inside extended, low excitation gas disks. These galaxies are markedly
different than local ULIRGs and high-z submm galaxies and quasars, where  higher
excitation and more compact gas is found.

\end{abstract}


\section{Introduction}

Over the last decade, deep and wide multiwavelength galaxy surveys have
been key in addressing critical issues regarding galaxy evolution.  By
using a variety of color selection techniques, and deriving
photometric and/or spectroscopic redshifts, different galaxy populations
have now been probed to unprecedented detail up to at least
$z\sim3$. Also, the history of cosmic star formation as well as
the corresponding build--up of stellar mass has now been constrained
using different observational approaches.

Less well understood are the physical processes that regulate the star
formation activity in these young systems and the question of if and how
galaxy star formation at high redshift differs from what is seen in
the local Universe. While it has been established that star formation
rates (SFRs) in galaxies were on average higher in the past and that
the cosmic SFR density peaked at $z>1$ (see, e.g., LeBorgne et al. 2009), 
it is still a matter of debate how much of
this increase is due to more frequent merging/interactions between
galaxies, or other processes. Analysis of the ultraviolet (UV) and
optical rest frame morphology and H$\alpha$ velocity fields of high redshift
galaxies suggest that the galaxies responsible for the bulk of the SFR
density at $z\sim1$~to~3 are disks (e.g., Bell et al.
2005; Elbaz et al. 2007; Genzel et al. 2008). These findings confirmed 
early hints for the presence of extended disks at high redshift, based
on damped Ly$\alpha$ studies (e.g., Wolfe et al. 1986).
Some authors,
 however, advocate a predominant role of mergers based on, e.g.,
 irregularities in high signal to noise (S/N) 3D observations of emission lines (e.g., Flores 
 et al. 2006). On the other hand, it is also found that distant
 star forming galaxies show irregular morphologies with high
 clumpiness, especially in the rest-frame UV (e.g., Cowie et al. 1996; Chapman et al. 2003;
 Daddi et al. 2004; Elmegreen \& Elmegreen 2005).
 In this context it is interesing to note that the
 existence of such bright clumps can explain the observed kinematic
 irregularities in high--z galaxies (i.e., without the need to require
 the presence of mergers, Immeli et al. 2004; Bournaud et al. 2008). More direct evidence
 for an important role of in situ, quiescent star formation in distant
 galaxies came from the estimate that the duty cycle of star formation
 is high in massive $z\sim2$ galaxies (Daddi et al. 2005; 2007a; see also Caputi et
 al. 2008), with typical durations of $\approx0.5$--1~Gyr. This is much longer than the typical duration
 expected for starbursts ($\simlt100$~Myr)  triggered by mergers,
 based on simulations (e.g., Mihos \& Hernquist 1996; Di~Matteo et al. 2008) and also on the observed gas consumption timescales
 for IR luminous merging galaxies (e.g., Downes \& Solomon 1998).

A recent major step forward in characterizing the nature of star
formation in distant galaxies was based on the discovery that star
forming galaxies define a narrow locus in the stellar mass-SFR plane
(in this paper we will refer to the ratio of SFR to stellar mass as
the specific SFR, SSFR).  This correlation has been estimated to have
a slope of 0.7--1.0 in log space and a dispersion of a factor of
only $\sim$2 from redshifts $z\sim0.1$ to $z\sim3.0$ (Noeske et
al. 2007; Elbaz et al. 2007; Daddi et al. 2007a; Magdis et al. 2009).

The normalization in this correlation, i.e. the SSFR at a given stellar mass,
is higher by a factor of 30 at
$z=2$--3 compared to the local Universe. At even higher redshifts the
situation is unclear as the SFRs are typically less well constrained
(as they are primarily based on UV luminosities). However, the
available data is still consistent with the picture that the
correlation extends to earlier epochs ($3.5<z<7$) with a normalization
similar to that seen at $z=2$ (Daddi et al. 2009; Stark et al. 2009,
Gonzales et al. 2009).  
The existence of this correlation has major implications for our
understanding of galaxy formation. For example, the cosmic evolution
of the normalization of the relation (Daddi et al. 2009; Pannella et
al. 2009; Gonzales et al. 2009) is similar to the cosmic evolution of
the SFRD (Hopkins \& Beacom 2006; 2008; Le Borgne et al. 2009, etc). It is
possible in fact to describe most of the SFRD evolution as the
combined result of the evolution of the stellar mass-SFR correlation
and that of the space densities (e.g., the mass/luminosity functions)
of galaxies.
The possible peak at $z\sim2.5$--3  (Pannella et al. 2009; Magdis et al. 2009;
Daddi et al. 2009a)
of the SSFRs is thus likely
more fundamentally related to the physics underlying galaxy formation
than the peak at $z>1$ of the SFRD evolution, as it is hinting at a
change in the physical processes that regulate star formation.

The remarkable tightness of this relation
implies a very high degree of homogeneity in these galaxies.
In other words, despite the physical complexity of the
process by which stars are formed, for a normal, 
near-IR selected star forming galaxy, the
star formation rate is ultimately linked to its stellar mass (which in
itself is presumably related to the mass of the underlying dark matter
halo).

   \begin{figure*}[!ht]
   \centering
   \includegraphics[width=13.0cm,angle=0]{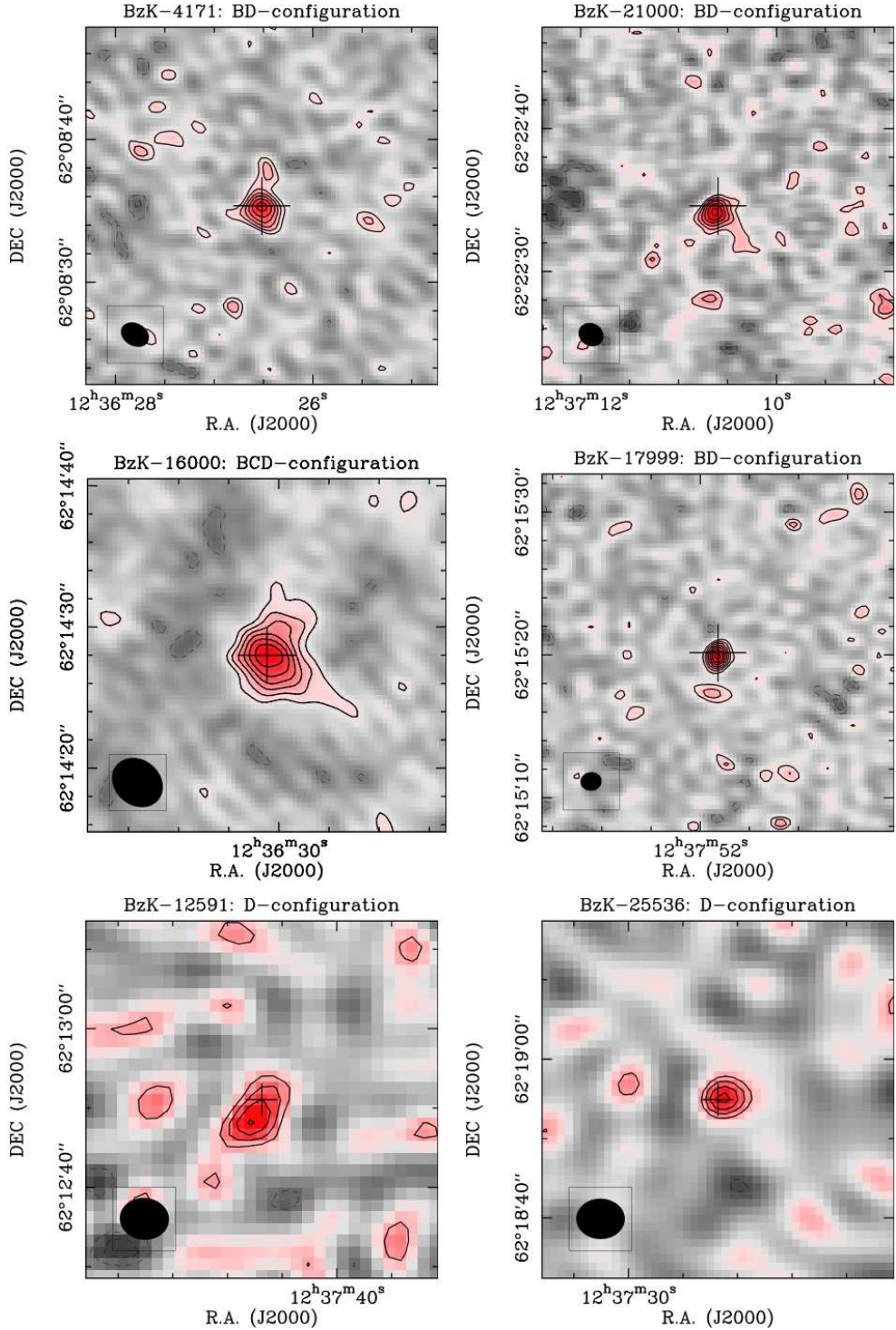}
   \caption{Velocity averaged maps of the CO[2-1]  emission
   in our sample BzK galaxies. The (cleaned) images were
   obtained using natural weighting and are based on data from all
   observations (as indicated on top of the individual
   panels). Contour levels start at $\pm2\sigma$ and are in steps
   of $1\sigma$ (see Tab.~\ref{tab:1}). The size
   and orientation of the beam is indicated in the bottom--left
   corner.  The panels in the top two rows are 25$''$ in size, those
   in the bottom row are 45$''$ in size. The cross in each panel corresponds
   to the VLA 1.4~GHz radio continuum position (see Tab.~\ref{tab:1})
   and are $\pm2''$ ($\pm 17$~kpc for $z=1.5$) in size.
           }
              \label{fig:2D}%
    \end{figure*}

   \begin{figure*}[!ht]
   \centering
   \includegraphics[width=13.0cm,angle=0]{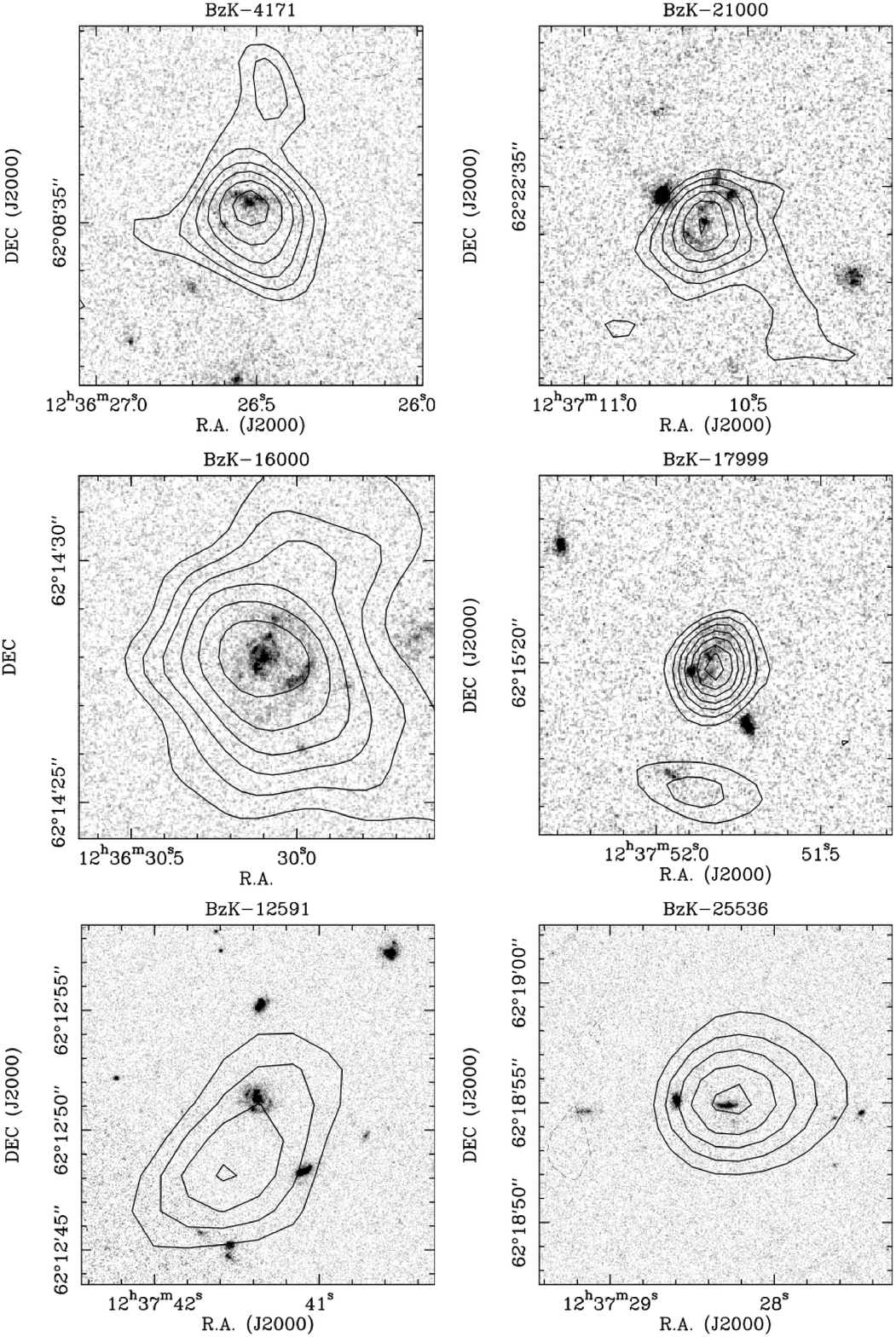}
   \caption{ACS images in the F775W band (i-band) of the target BzK galaxies. CO[2-1]  contours are the same as shown
in Fig.\ref{fig:2D}. The panels in the top two rows are $7.5''$ in size.
The panels in the bottom are 15$"$  in size. For reference, 
1$''$ subtends 8.5~kpc at $z=1.5$.
           }
              \label{fig:acs}
    \end{figure*}

   \begin{figure*}
   \centering
   \includegraphics[width=13.0cm,angle=0]{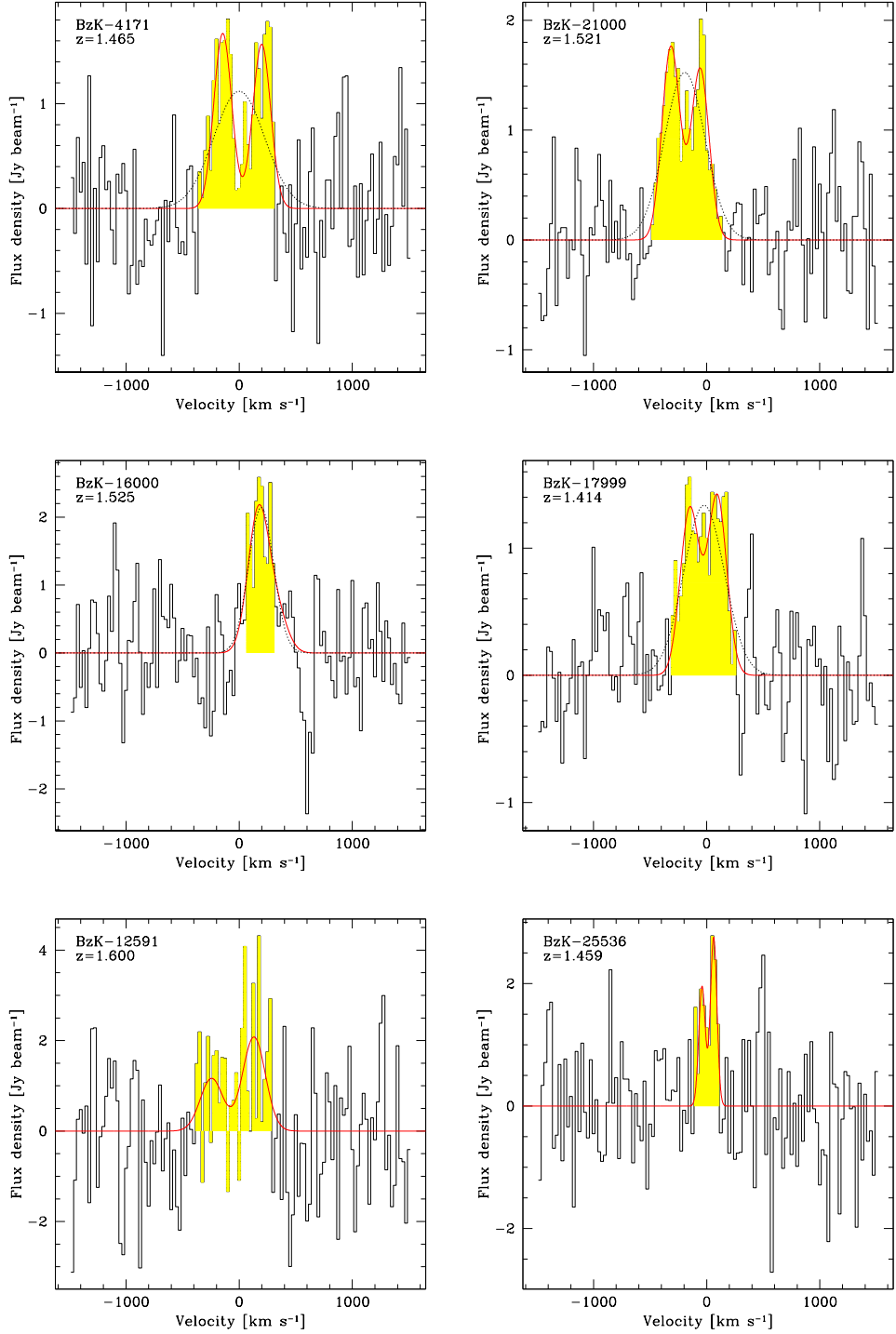}
   \caption{Spectra of CO[2-1] emission binned in steps of 25~km~s$^{-1}$. The yellow color indicates the regions where 
   positive emission is detected. These regions have been used to derive total
   integrated fluxes and the velocity averaged
   maps in Fig.~\ref{fig:2D}. The red lines show best fitting double
   Gaussian profiles to the spectra (see Tab.~\ref{tab:2} for fit values). In
   the fits, we forced the FWHM
   of the two Gaussian functions to be the same within a given galaxy. For
   the four galaxies with the highest quality spectra we also show
   single Gaussian fits. Those yield significantly worse fits to the
   wings of the CO profiles, except for the face-on galaxy BzK-16000 that is consistent with a single Gaussian.
   Velocities are computed with respect to the tuning frequencies
   of the observations as listed in Tab.~\ref{tab:1}.
           }
              \label{fig:1D}
    \end{figure*}

   \begin{figure*}
   \centering
   \includegraphics[width=17cm,angle=0]{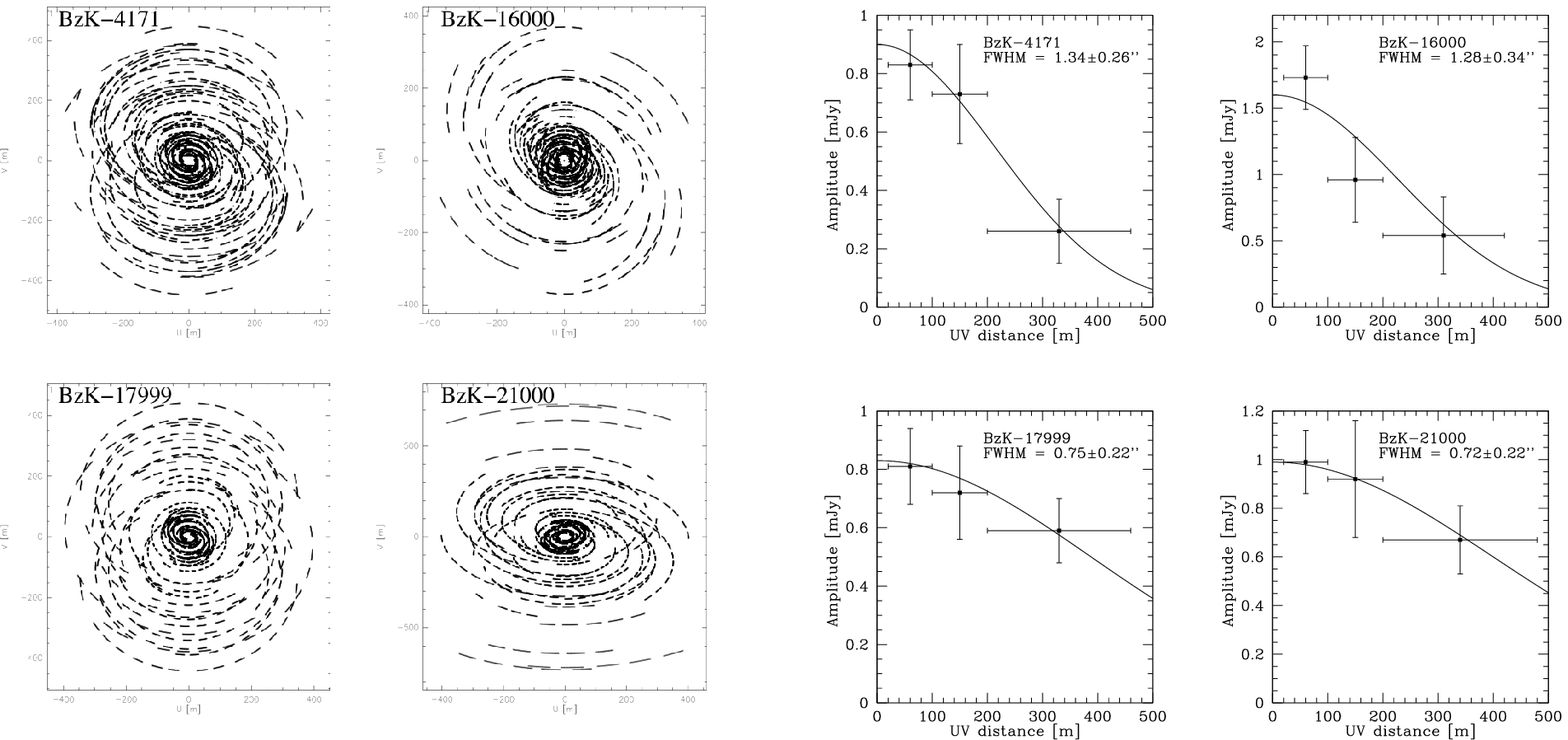}
   \caption{{Left:} uv-coverage of the four target BzK galaxies that were observed also with the higher resolution B-configuration. {Right:} Signal amplitude versus
   baseline length in the uv plane. The solid curves show circular
   Gaussian models with FWHM as labeled. As a reminder, 
   for a point/unresolved source the
   visibility amplitudes would be constant with uv distance.
           }
              \label{fig:uvcov}
    \end{figure*}

In order to make further progress in understanding how galaxy
formation proceeds in distant galaxies, and to shed light on the
origin of the stellar mass-SFR correlations, we need to start probing
the properties of molecular gas reservoirs in these galaxies, since it is
the molecular gas out of which stars form.
If the assembly of galaxies is mostly driven by mergers, 
one would expect most star formation in merger-driven starbursts 
with concentrated
gas cores (like in submm galaxies) and very efficient star formation, and much less molecular gas and star formation in regular
objects. If, on the other hand, most star formation is due to smooth gas flows,
then 
star formation would predominantly
take place in more quiescent disk galaxies, accreting gas from their 
surrounding medium.
In addition, one could
speculate that the tightness in the stellar mass-SFR correlation might
be pointing to homogeneous properties of the molecular gas reservoirs and star formation modes
inside the galaxy populations.  The increase in the normalization of
this correlation with cosmic lookback time could then possibly be simply
linked to an increase of the average gas fractions in galaxies, as well as in the ratio of the molecular (H$_2$) 
over the atomic (HI) hydrogen fraction (see, e.g., Obreschkow \& Rawlings 2009).  The
presence of massive gas reservoirs seem to be a major requirement for
justifying high SFRs, with the lack thereof possibly supporting
instead the need for changes in the initial stellar mass function (IMF).

Recently, we reported the detection of large molecular gas reservoirs
in two near-IR selected massive galaxies at $z\sim1.5$ (Daddi et al.\
2008). Based on the knowledge available at the time, i.e., the
previously known empirical  correlation between CO line luminosity ($L'_{\rm
CO}$) and total IR luminosity ($L_{\rm IR}$) those galaxies should not
have been detected in CO. However, observations were still attempted
based on the results of Daddi et al. (2005; 2007) that suggested a
high duty cycle and long duration for star formation in massive
high-$z$ galaxies. The subsequent detections of CO[2-1] emission
indeed implied low star formation efficiencies
($SFE$, defined as the ratio $L_{\rm IR}/ L'_{\rm CO}$). These SFEs
are similar to what is seen in local spirals and much lower than
previously seen in high redshift galaxies. Subsequent studies revealed
low-excitation CO [3-2] to [2-1] line ratios in these
galaxies (Dannerbauer et al. 2009) that are more typical of local spirals.

Given that both galaxies were detected in CO, this early result
suggested commonplace large molecular gas reservoirs in distant massive
galaxies (Daddi et al. 2008). However, improved statistics
would clearly be needed in order to formally confirm the result and to
estimate the variations in gas properties among the high-$z$ galaxy
populations. Higher spatial resolution observations are further
required to constrain the sizes of the molecular emission; i.e., given
the low SFEs one
would expect extended reservoirs rather than the compact emission lines seen in
submm selected galaxies (SMGs) for which Tacconi et al. (2006; 2008) find
typically half light radii of order 2~kpc.

In this paper we report on low-resolution (D-configuration)
observations with the IRAM Plateau de Bure Interferometer (PdBI) of
CO[2-1] (redshifted to 3mm) in four
 additional BzKs (near-IR selected galaxies at
$z\sim1.5$) thus tripling the number of sources in our sample. In
addition, we report on higher resolution ($\sim1$--1.5$''$; B-configuration) 
observations of four galaxies. We also
discuss the properties of these six galaxies with respect to
the `parent' BzK sample. We use the information gathered on these six
galaxies in order to independently constrain the molecular gas mass
and, conversely, the infamous CO luminosity to gas mass
 conversion factor $\alpha_{CO}$.

The paper is organized as follows. In section~\ref{sec:data} we present
the new PdBI observations and their analysis.  
In Sect.~\ref{sec:sample} we discuss the relation of the six galaxies to
the parent sample and the derivation of accurate SFRs and stellar masses.
The SFEs are
characterized in Sect.~\ref{sec:LCO} and compared to other galaxy
populations. We introduce numerical models of clumpy disks in
Sect.~\ref{sec:models} and use the models to calibrate dynamical
masses for our sample in Sect.~\ref{sec:Mdyn}. This section also
discusses implications for the implied gas masses and the conversion
factors.  A discussion and summary of our results is presented in
Sect.~\ref{sec:discussion}~and~\ref{sec:summary}. 
We apply a concordance WMAP3 cosmology throughout the paper.

\section{PdBI observations and results}
\label{sec:data}

\subsection{Target selection}
\label{sec:targets}

The six galaxies that were observed in CO were culled from the sample of
Daddi et al. (2007ab) in the Great Observatories Origins Deep Survey North field 
(GOODS-N).  This sample is based on
galaxies selected in the K-band down to $K<20.5$ (Vega scale; or
$K<22.37$ AB), from which we selected the population of star forming
galaxies at $1.4<z<2.5$ by using the "star forming" BzK color
criterion of Daddi et al. (2004b) and requiring a detection in deep
Spitzer imaging at $24\mu$m. From these galaxies, we chose those
that had a pre-determined spectroscopic redshift as required for
follow--up observations in CO. Currently, 89 BzK galaxies of the Daddi et al. (2007) sample
in GOODS-N have a spectroscopic redshift, from various observing campaigns and  
from the literature. 
Among the six targeted galaxies, five redshifts were obtained through our 
GOODS-N campaigns at Keck using DEIMOS (Stern et al. in
preparation). The redshift for BzK-12591 was derived by Cowie et
al. (2004). This galaxy, having a strong bulge evident from HST
imaging, has a possible detection of [NeV]$\lambda$3426\AA\ emission
line, suggesting the presence of an AGN.  This galaxy is not
detected in hard X-ray emission in Alexander et al. (2003) but exhibits some
soft X-ray emission.  In addition to the availability of a
spectroscopic redshift, we also required the detection of the galaxy
at 1.4~GHz. This is important because the radio continuum allows us to
derive an independent estimate of the SFR in the sources.

\subsection{Observations and data reductions}

Tab.~\ref{tab:1} summarizes the PdBI observations of the 
six massive, near-IR selected BzK galaxies at redshifts $1.4<z<1.6$
in the GOODS North field
(D-configuration data for two galaxies were already presented in
Daddi et al. 2008).  
The frequencies during the observations
were tuned to the expected redshifted frequency of CO[2-1].

\begin{table*}
{\footnotesize
\caption{Observation summary table}         
\label{tab:1}      
\centering                          
\begin{tabular}{l c c c c c c c c}        
\hline\hline                 
Source & RA$^1$ & DEC$^1$ & Conf. & Obs.dates & T$_{int}$$^2$ & Frequency & Combined beam$^3$ & rms \\

 & J2000 & J2000 &  &  & h & GHz & &  $\mu$Jy$^4$\\
\hline                        
BzK-4171 & 12:36:26.516 & 62:08:35.35   & D & 04-08/2007 & 8.2 & 93.525 & 2.05$''\times1.55''$, $PA=64^o$ & 41 \\     
          & &  & B & 01-03/2008  & 11.2 & &  \\      
\hline
BzK-21000 & 12:37:10.597 & 62:22:34.60  & D & 04/2007  & 6.3 & 91.375 &  $1.87'' \times 1.60'', PA=64^o$ & 51  \\      
          & &  & B  & 01/2008  & 6.2 & &  \\      
\hline
BzK-16000 & 12:36:30.120 & 62:14:28.00   & D & 05/2008  & 3.3 & 91.375 & $3.83'' \times 3.22'', PA=47^o$ & 54 \\      
           & & & C & 11/2008  & 2.8$^5$ & & \\      
           & & & B & 01/2009  & 4.4 & &  \\      
           & & & D & 05-07/2009  & 6.5$^6$ & &  \\     
\hline
BzK-17999 & 12:37:51.819 & 62:15:20.16   & D  & 05/2008   & 6.0  & 95.501 & $1.52'' \times 1.30'', PA=96^o$ & 41 \\ 
          & &  & B  & 01/2009 & 6.4  & & \\      
\hline
BzK-12591 & 12:37:41.371 & 62:12:51.06   & D  & 05/2008   & 5.4 & 88.669 & $6.37'' \times 5.26'', PA=85^o$ & 116  \\ 
\hline
BzK-25536 & 12:37:28.357 & 62:18:54.91   & D  & 05/2008  & 6.2 & 93.754 & $6.32'' \times 5.10'', PA=90^o$ & 72 \\ 
\hline                                   
\end{tabular}\\
Notes:\\
1: coordinates are from VLA 1.4~GHz continuum emission (Morrison et al. 2010). The VLA 1.4~GHz map has a resolution 
of 1.8$''$, and the typical position accuracy for our sources is $0.20''$\\
2: equivalent six-antennas on source integration time. \\
3: beam resulting after combining all available configuration and imaging with natural weighting. These are the
beams sizes displayed in Fig.~\ref{fig:2D}\\
4: noise per beam averaged over the full 1~GHz spectral range\\
5: data taken with a primary beam attenuation of 24\% while observing the source GN10 (Daddi et al. 2009b)\\
6: data taken with a primary beam attenuation of 40\% while observing another source \\
}
\end{table*}
We used the compact D-configuration of the antennas for detection of
the sources, providing the highest sensitivity and lowest spatial
resolution ($\sim$5.5$''$ at our typical frequency of 90~GHz).  Four
of the sources were followed up at higher resolution ($\sim$1.0--1.5$''$)
using the more extended B-configuration. 
Typical integration times were 5--10h per
configuration. Most of the observations were obtained with six antennas,
with the exception of a few tracks in which only five antennas were
available.  The correlator has 8 independent units, each covering
320~MHz (128 channels each with a width of 2.5~MHz) with a single
polarization, resulting in a total bandwidth of about 1~GHz with both
polarizations.

We reduced the data with the GILDAS software packages CLIC and MAP,
similarly to what is described in Daddi et al.\ (2008; 2009ab) and
Dannerbauer et al. (2009).  Bandpass calibration was performed using
observations of the standard calibrators J0418$+$380 and 3C273.
Observations of J1044$+$719 and J1150$+$497 were used for phase
and amplitude calibration.  For flux calibration we used MWC~349
as a primary calibrator and 3C273 when the former was not available.
A model was used to account for the frequency dependence and time
variability of the emission of MWC~349. Given the typical observed scatter
between the observations and the model for MWC~349, and also comparing
the results of flux calibration to the implied antennas efficiency
variations, we estimate that the flux calibration is accurate to 10\%
at 90~GHz.

\begin{table*}
{\footnotesize
\caption{Observed properties}            
\label{tab:2}      
\centering                          
\begin{tabular}{l c c c c c c c c c}        
\hline\hline                 
Source & $z_{\rm Keck}^1$ & $z_{\rm CO}$ & $v_{\rm CO}$ FWHM$^2$ & $I_{\rm CO[2-1]}$~$^2$ & S/N det$^3$ &
CO size & CO blue/red$^{4}$ & PA blue/red & S(1.4~GHz) \\
 & & & km~s$^{-1}$ & Jy km s$^{-1}$ &  & kpc & kpc & & $\mu$Jy \\
\hline
BzK-4171 & 1.465 & 1.46520 $\pm$ 0.00058 & 530 $\pm$ 32 & 0.65 & 8.3 & 11.3 $\pm$ 2.2 & $1.7\pm1.7$  & -- & 34 \\
BzK-21000 & 1.523 & 1.52133 $\pm$ 0.00036 & 444 $\pm$ 26 & 0.64 & 9.4 &  6.2
$\pm$ 1.9 & $5.3\pm1.3$ & $-25\pm13^{\rm o}$ & 43  \\
BzK-16000 & 1.522 & 1.52496 $\pm$ 0.00028 & 194 $\pm$ 18 & 0.46 & 8.7 &  10.9 $\pm$ 2.9 & --  &-- & 19 \\
BzK-17999 & 1.414 & 1.41385 $\pm$ 0.00032 & 440 $\pm$ 33 & 0.57 & 9.4 & 6.4 $\pm$
1.9 & $5.7\pm1.3$ & $-1\pm12^{\rm o}$ & 34  \\
BzK-12591 & 1.600 & 1.60019 $\pm$ 0.00076 & $\sim600$ & 0.84 & 5.3 &  --  & --  &-- & 186 \\
BzK-25536 & 1.459 & 1.45911 $\pm$ 0.00013 & $\sim170$ & 0.36 & 6.4 &   -- & --  &-- & 28 \\
\hline                                   
\end{tabular}\\
Notes:\\
1: the accuracy of Keck redshift is generally better than 0.001, except for BzK-16000 and 21000 (see Section~\ref{sec:spectra})\\
2: derived from fitting double Gaussian profiles\\
3: to determine the S/N of the detection we add in quadrature the S/N from point source fitting of the data
in each configuration, using the spectral regions with positive line emission (yellow regions in
Fig.~\ref{fig:1D}) \\
4: spatial separation between the red and blue component in the CO[2-1] spectra
}
\end{table*}

\begin{table}
\caption{Rest-frame UV morphology}             
\label{tab:3}      
\centering                          
\begin{tabular}{l c c c c c}        
\hline\hline                 
Source & r$_{\rm e}$ & n S\'ersic & a/b & PA & FWHM$_{\rm circ}$ \\
 & kpc & &  & & kpc \\
\hline
BzK-4171  & 3.7 & 1.0 & 0.40 & 86$^{\rm o}$ & 3.4    \\
BzK-21000 & 6.2 & 0.5 & 0.37 & $-35$$^{\rm o}$ & 11.0    \\
BzK-16000 & 5.1 & 1.5 & 0.97 & -- & 5.9    \\
BzK-17999 & 4.7 & 0.5 & 0.42 & $-39^{\rm o}$ & 7.7    \\
BzK-12591 & 4.5 & 2.0 & 0.85 & 32$^{\rm o}$ & 6.0    \\
BzK-25536 & 3.0 & 1.0 & 0.32 & 90$^{\rm o}$ & 3.4    \\
\hline                                   
\end{tabular}\\
Notes:\\
The formal {\em Galfit} errors on the sizes and axis ratios correspond to an accuracy of better than 5\%.
The PA is measured East of North.
FWHM$_{\rm circ}$ is derived from the fit of a circular Gaussian profile.
\end{table}

CO[2-1] emission has been successfully detected at the expected
frequency for all six galaxies.
Fig.~\ref{fig:2D} shows the detection images (natural weighted,
cleaned) averaged over the spectral channels where emission was
detected. In all cases the peak flux is detected at $S/N>5$.
Fig.~\ref{fig:acs} show the same CO[2-1]
contours overlayed on the Hubble Space Telescope (HST) images of the
galaxies from the GOODS Advanced Camera for Surveys (ACS) campaign (Giavalisco et al. 2004).  We note
that in five of the six cases, the CO[2-1] is very well centered on the VLA
1.4\,GHz radio continuum positions (Morrison et al. 2010; crosses in
Fig.~\ref{fig:1D}) and on the ACS rest frame ultraviolet images
(Fig.~\ref{fig:acs}). However, for object BzK-12591 the CO[2-1]
detection is offset by about half a beam or
$\sim3''$. The CO[2-1] detection for this galaxy is the one with the
lowest S/N ratio and is based only on low--resolution D-configuration
data. We conclude that the offset is most likely due to a combination
of the effects of noise (we expect uncertainties of 0.6--1$''$ at $1\sigma$ in the position) 
and phase instabilities (the interferometric seeing
was about 1.9$''$) during the late-spring
observations (Tab.~\ref{tab:1}).

\subsection{CO[2-1] detection images and spectra}
\label{sec:spectra}

Fig.~\ref{fig:1D} shows the resulting CO[2-1] spectra binned with a
resolution of 25~km\,s$^{-1}$. The spectra were obtained by fitting the data in uv
space with the following procedure: we first averaged the uv data over
the spectral range in which CO has been detected (see the yellow
shaded regions in Fig.~\ref{fig:1D}) to determine the spatial centroid
of the CO emission. Keeping this centroid fixed for all channels, we
then extracted the spectra with 25~km~s$^{-1}$ binning by fitting each
channel in the uv data. For those objects for which we also have
B-configuration data and that appear resolved (BzK-4171, 21000, 16000
and 1799, see next section) we used a circular Gaussian model with a
fixed FWHM (derived from the velocity averaged data).  For the
remaining two targets (BzK-12591 and 25536) we fitted point source
models.

To determine the properties of the CO[2-1] emission (observed flux,
velocity centroid and FWHM) we fitted Gaussian functions to the
observed spectra (Fig.~\ref{fig:1D}).  Several of the spectra show
clear evidence for double-peaked emission profiles, suggestive of
rotation. We obtained fits for all galaxies with a single Gaussian and
with two Gaussian functions (in this case fixing the FWHM in each component
to the same value).  In most cases we find that a fit with two Gaussian
profiles provides a better description of the data.  This
includes objects like BzK-21000 and 17999 where, despite the fact that
the evidence for a double peak profile is only tentative, a fit with a
single Gaussian is disfavored as it is not able to accurately
reproduce the rapid decrease of the flux at the edges of the spectra.
In no case we do find evidence for continuum emission. We have combined all 
line-free channels in all galaxies to derive an
average continuum at 3.3mm of $<100\mu$Jy at the 3$\sigma$ level.
This is consistent with
the expectation, given the moderate SFRs in these galaxies (see Section~\ref{sec:sample}), 
that the continuum emission should be at the level of 20--30$\mu$Jy.

   \begin{figure}
   \centering
   \includegraphics[width=8.8cm,angle=0]{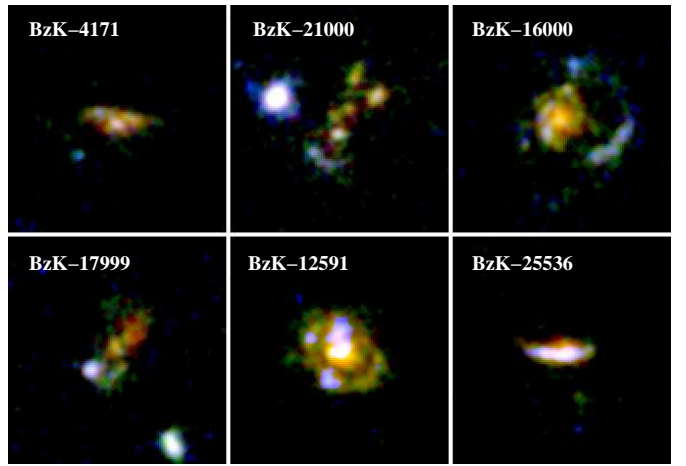}
   \caption{3-color RGB imaging of the six CO detected BzK galaxies, based on ACS imaging
   of Giavalisco et al. (2004), release 2.0. Red is F850LP, green is F775W, blue is F435W.
   We have used a 2-pixel tophat smoothing to improve the source visibility. Images are $3''$ in size.
           }
              \label{fig:3color}
    \end{figure}

   \begin{figure*}
   \centering
   \includegraphics[width=17.0cm,angle=0]{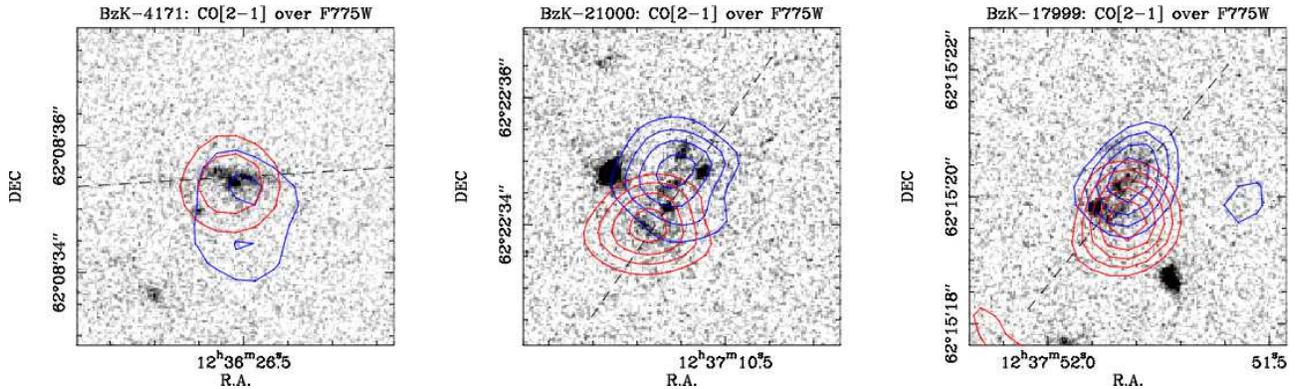}
   \caption{Contours of CO[2-1] are overlayed, separately for the blue (approaching) 
   and red (receding) velocity ranges in the spectra,
   on ACS F775W images ($5''$ size).  Contours start at
   $\pm2\sigma$ and increase in steps of 1$\sigma$. 
   The displacement of the red and blue parts of CO[2-1] for BzK-21000
   and 17999 appear to be well aligned with the
   rest-frame UV major axis (dashed lines).
           }
              \label{fig:BzK_ALL_rotation}
    \end{figure*}

Tab.~\ref{tab:2} summarizes the results of the spectral
fitting. Typical integrated CO[2-1]  fluxes are
0.4--0.8~Jy~km~s$^{-1}$. The full width at half maximum (FWHM) of the
CO[2-1] emission is typically in the range of 400-600~km~s$^{-1}$,
with two notable exceptions: BzK-16000 and 25536. For BzK-16000 the
narrower velocity range is most likely due to the fact that the galaxy
is observed nearly face-on. For
BzK-25536 this is likely, at least in part, due to the relatively low S/N ratio of
this source (see later in Sect.~\ref{sec:Mdyn}).  The
velocity centroid of the CO[2-1] emission can be used to derive a
CO[2-1] redshift for the six galaxies.  The CO redshifts agree well      
with the Keck redshifts to within $\Delta z= (z_{\rm CO}-z_{\rm Keck})=0.001$, the typical
accuracies of the latter.  Two objects show a larger offset: BzK-16000
has $\Delta z = 0.003$ that can be, at least
partly, explained due to the lower accuracy of the Keck redshift. In this case, the
optical redshift was determined through weak absorption lines in the
UV rest frame (all other redshifts are from the detection of
[OII]$\lambda3727$ emission lines); BzK-21000 has $\Delta z=-0.0017$,
but its [OII] emission line is close to a very bright, telluric OH emission line.

\subsection{Estimation of the spatial size of CO[2-1]}

When comparing the D-configuration and B-configuration data 
(their respective uv coverages are shown in Fig.~\ref{fig:uvcov}), imaged
separately, we find brighter peak flux densities in the
D-configuration only data. This already indicates that the sources are
spatially resolved.  In order to quantitatively confirm this finding
and to estimate the intrinsic size of the CO[2-1] emission we have
fitted models to the uv data. We have averaged the data over the
spectral ranges showing positive emission (yellow-shaded regions in
Fig.~\ref{fig:1D}).  Given the modest S/N
ratio of the observations, we fitted circular Gaussian models to the
data to limit the number of free parameters (four in this case: the
spatial center RA-DEC positions, the total flux and the Gaussian
FWHM).  Fig.~\ref{fig:uvcov} (right) shows the total CO[2-1] 
amplitude binned in ranges of uv distances, covering baselines from
$\sim50$m to $\sim300$-400m.  The overplotted model corresponds to the
expected behavior in uv space of circular Gaussian profiles with the
fitted sizes. In all four cases we find that the sources are spatially
resolved with a significance of $>3\sigma$ ($5\sigma$ for
BzK-4171). The FWHMs of the circular Gaussian functions range within
0.7--1.3$''$, corresponding to 6--11~kpc at the redshifts of the
galaxies.  

\subsection{Comparison to ACS imaging}

It is interesting to compare the CO[2-1] sizes to those from the HST
imaging. In order to do that, we used the {\em Galfit} code (Peng et
al. 2002) to characterize the ACS z-band images (F850LP filter),
corresponding roughly to U-band rest frame. Tab.~\ref{tab:3}
summarizes the results of the S\'ersic profile fitting. Preliminary
results for the S\'ersic profile fitting of BzK-21000 and 4171 were
presented also in Daddi et al. (2008), and are consistent with the new
measurements based on HST images that have twice the integration
time (ACS GOODS release 2.0). Most of the sources are consistent with
$n_{\rm Sersic}\simlt1.5$ profiles that are typical for spiral disks.
For BzK-12591 and 16000 we find higher S\'ersic indexes of 1.5 and
2. Inspection of the HST images (Fig.~\ref{fig:3color}) shows that
these galaxies appear to be disk galaxies with a significant bulge
component. 
We cannot exclude that bulges might exist in some of the other
galaxies as well, and deep high resolution near-IR observations (e.g., with WFC3 on HST) would be
necessary to check this.  Two galaxies have a S\'ersic index lower
than 1: BzK-17999 and 21000. For these two galaxies most of the UV
light is concentrated in bright clumps, implying a light distribution
flatter than an exponential. This is consistent with the appearance
expected for gas rich disks (Bournaud et al. 2008).
We notice that in most cases the target BzK galaxies display a
significant inclination in the ACS imaging, with axis ratios often
below 0.5, i.e. they are not circularly symmetric. This implies that the size estimate derived from S\'ersic
model fitting might not be directly comparable with the cruder
CO[2-1] sizes derived from (symmetric) circular Gaussian fitting. 
While it is not meaningful to
fit our CO[2-1] observations with more complex spatial shapes due to
the limited S/N ratios, we decided to fit also the ACS images  with circular
Gaussian models to test for consistency between CO[2-1] and rest
frame UV sizes.

The results of this analysis are listed in Tab.~\ref{tab:3}.
For three galaxies the size
measurements agree within 1-2$\sigma$. 
The exception is BzK-4171 for
which the CO[2-1] size estimate is about three times larger than the
UV rest frame size, a difference significant at the 3.4$\sigma$ level.
This is also the object with the most accurate CO[2-1] size
determination in our sample.  If we average over the sample in order
to increase the S/N ratio, we find that the ratio of CO[2-1] to
rest-frame UV sizes is 1.24 for all the four galaxies and 0.96 when
excluding BzK-4171. Therefore, it appears that CO[2-1] and
rest-frame UV sizes are quite similar, within 20\% 
on average.

   \begin{figure}
   \centering
   \includegraphics[width=8.8cm,angle=0]{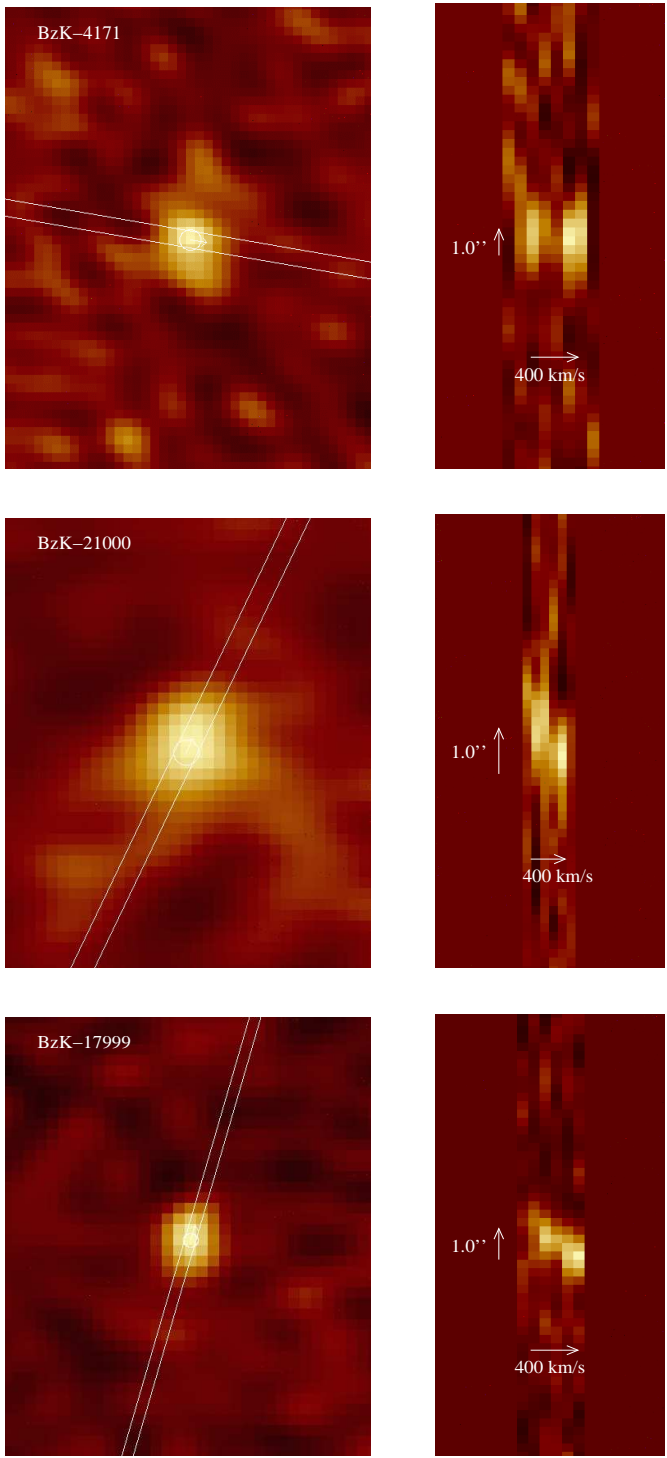}
   \caption{Left panels: CO[2-1] images with superimposed pseudoslits that are used to extract position--velocity
   diagrams.
   Right panels: the respective position velocity diagrams. The arrow shows the spatial and velocity scales in the
images (1$''$ corresponds to 8.5~kpc at $z=1.5$) .
   The data were binned in velocity (width: 100~km~s$^{-1}$).
           }
              \label{fig:BzK_ALL_PV_fig}
    \end{figure}

\subsection{Evidence for rotation}

We have analyzed the high--resolution CO[2-1] data of three of our
galaxies (as one, BzK-16000, appears to be seen face--on) to check for
signatures of rotation.  Solid evidence for galactic rotation in
near-IR selected massive BzK galaxies has recently been derived using
H$\alpha$ observations from the SINS survey (e.g., F\"orster-Schreiber et
al. 2009). Clearly, it would be important to confirm such
findings based on the dynamics of the molecular gas that should trace
the underlying velocity field.

In Fig.~\ref{fig:BzK_ALL_rotation} we show CO[2-1] contours imaged
separately for the blue-- and red--wing velocity ranges of BzK-4171,
21000 and 17999. For BzK-21000 and 17999 we observe a significant
spatial displacement of the red (receding) versus blue (approaching)
component. Their displacement is, not unexpectedly, aligned with the
galaxy major axis as determined from the UV rest frame
observations. Also, the separation of the blue and red components is
found to be quite consistent with the overall UV sizes of the galaxies.
We conclude that
we find evidence for galactic rotation in the disks in at least
a couple of our sources.

For a more direct assessment we obtained position-velocity diagrams
for our galaxies (Fig.~\ref{fig:BzK_ALL_PV_fig}). In order to obtain
sufficient S/N we have binned the data over 100~km~s$^{-1}$ and
extracted position-velocity diagrams using pseudo-slits at the
orientation defined by the major axis in the ACS
imaging. As can be seen from Fig.~\ref{fig:BzK_ALL_PV_fig}, 
we again find evidence for rotation. In particular, BzK-21000
shows the expected S--shaped structure indicative of disk rotation,
consistent with a flat rotation curve.  A
less pronounced S-shaped rotation structure is also seen in BzK-17999.

In the case of BzK-4171 the approaching and receding parts of the
emission are not significantly offset
(Fig.~\ref{fig:BzK_ALL_rotation}), and no clear evidence for rotation
is present in the position-velocity diagram.  We explored position
velocity extractions at different orientations but could not find
evidence for rotation at any position angle.  We recall that for
BzK-4171 we also got the puzzling result that the CO[2-1] size was
found to be substantially larger than the rest-frame UV size. This
object has the most pronounced double horn profile in our sample, with
a very significant central minimum reaching about 15\% of the peak
flux of the two symmetric horns. 
This galaxy is the most compact
object in our sample in the rest-frame UV, and the half light radius
from the S\'ersic fit is consistent within the limits with the spatial
separation between the red and blue component. Higher resolution and higher S/N observations are required to
constrain the rotation in the disk of BzK-4171.

\section{General properties of the sample}
\label{sec:sample}

In this section we discuss the overall properties of the CO-detected
galaxies. We compare their colors, stellar mass and star formation
rates to the full sample of near-IR selected galaxies at similar
redshifts to investigate whether they are representative galaxies for
their mass. We further discuss the accurate derivation of the
star-formation rates through multiwavelength indicators. We also
explain the measurements of the stellar mass and stellar population
parameters through fitting of synthetic models to the observed
multicolor spectral energy distribution (SED).

\subsection{Comparison to the parent sample}

   \begin{figure}
   \centering
   \includegraphics[width=8.8cm,angle=0]{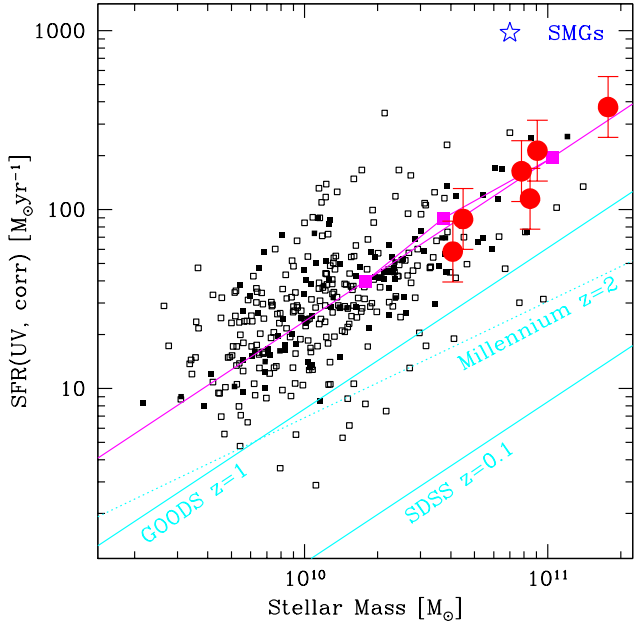}
   \caption{The location of the six BzK galaxies (large
   filled circles) in the stellar mass-SFR diagram.
   SFRs are derived from the UV luminosities at 1500\AA, corrected for
   dust reddening. We show typical error
   bars in the SFRs of the CO detected galaxies based on the results
   in Sect.\ref{sec:sample}.
   The smaller points (squares; filled symbols are for spectroscopic redshifts, which are 30\% of the sample) 
   show individual BzK galaxies from the GOODS South sample of Daddi et al. (2007ab)
 to probe smaller
   stellar masses (only star forming BzK galaxies, or sBzKs, are shown). The
   magenta line shows the derived correlation
   between stellar mass
   and SFR. Filled magenta squares show the SFRs measured from radio
   stacking as a function of stellar mass
   (see also Pannella et al. 2009). The
   light blue lines with lower normalization show
   the same correlation at $z=0.1$ and $z=1$ (Elbaz et al. 2007) and in the Millennium simulations at $z=2$
   (dotted line).
   Figure adapted from Daddi et al. (2007), but both SFRs and stellar
   masses were rescaled using a Chabrier IMF.
           }
              \label{fig:SFR_Mass_CO}
    \end{figure}

We have checked whether our source selection criteria (Sect.~\ref{sec:targets}) imposed
substantial biases on the choice of CO targets.
Fig.~\ref{fig:SFR_Mass_CO} shows the stellar mass-SFR diagram for the
CO--detected galaxies in comparison with the full sample of BzK
galaxies in GOODS-South. In the following we use the GOODS-South sample
for comparison even if the CO--detected galaxies are in the GOODS-N
field as our GOODS-South catalog reaches 1.5 mag deeper in K, down to
$K<22$ Vega. In order to use the same estimators for all galaxies in this plot, we use the
empirical relations by Daddi et al. (2004) to estimate stellar masses
and SFRs from UV luminosities in all objects (we discuss more
accurate derivations of these quantities for the CO--detected galaxies
below). From this plot it appears that the targeted galaxies 
are massive, but that they lie right on
top of the stellar mass-SFR correlation. The reason why they are
massive systems is due to the fact that we required them to
have a radio continuum detection: the
typical $5\sigma$ limits of 20$\mu$Jy in our 1.4~GHz VLA maps
(Morrison et al. 2010) correspond to SFRs of
$66$~M$_\odot$~yr$^{-1}$ at $z=1.4$ and $SFR=156$~M$_\odot$~yr$^{-1}$ at
$z=2.0$.

We note that the six galaxies in our sample are all at $1.4<z<1.6$;
i.e., in the lower redshift range probed by the BzK selection that
extends over $1.4<z<2.5$. This is due to the requirement of a
radio detection coupled with the choice to observe the 2-1 transition
of CO: such a transition has a rest-frame frequency of 230.538~GHz and
can be observed with the PdBI only up to $z=1.87$. We
have chosen to observe the 2-1 transition rather than higher
transitions in order to limit the uncertainties and corrections due to
the possible presence of low excitation molecular gas. Based on the
results by Dannerbauer et al. (2009), we would have underestimated the
total CO luminosity by a factor of $\sim2.5$ if we had observed the
CO[3-2] transition and assumed constant brightness temperature in the
CO emission. 
Observing CO[3-2] would have also resulted in substantially lower S/N ratio
for the galaxies in a fixed integration time, a finding consistent with the non detection of similar
galaxies in CO[3-2] by Tacconi et al. (2008) and Hatsukade et al. (2009).

   \begin{figure}
   \centering
   \includegraphics[width=8.8cm,angle=0]{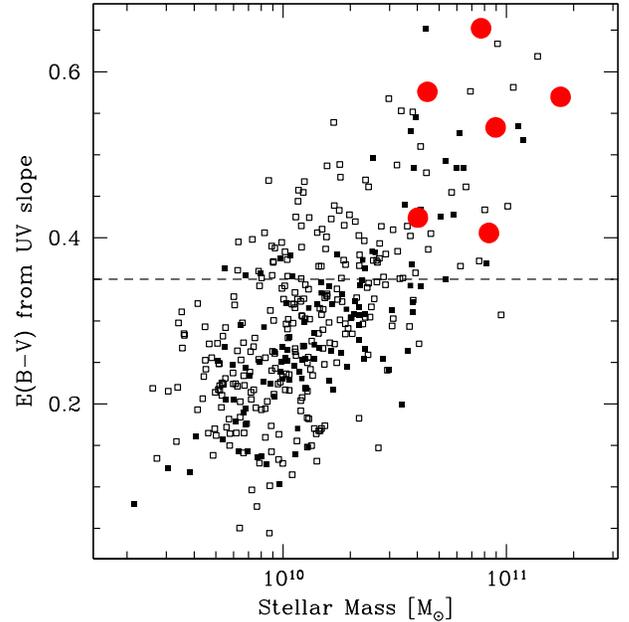}
   \caption{The reddening $E(B-V)$ estimated from the UV slope is plotted versus the stellar mass.
   The large filled circles are the six CO detected galaxies, while smaller symbols refer to the sample of near-IR
   selected BzK galaxies (only star forming BzK galaxies, or sBzKs, are shown). Objects above the horizontal dashed line are not
 selected with the BM/BX technique
   (Adelberger et al. 2004; Steidel et al. 2004).
           }
              \label{fig:EBV_Mass_CO}
    \end{figure}

   \begin{figure*}
   \centering
   \includegraphics[width=18.0cm,angle=0]{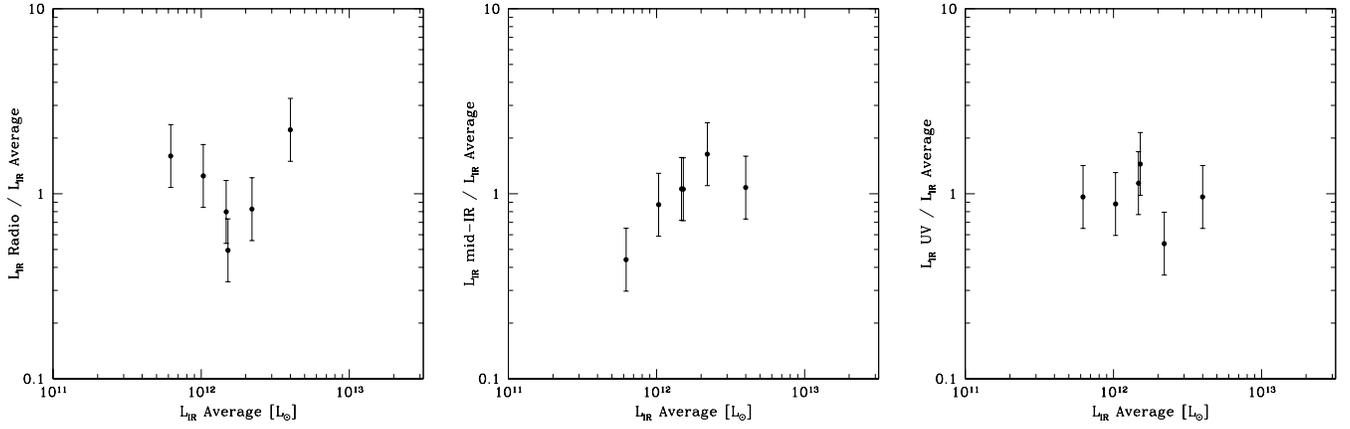}
   \caption{The IR luminosities ($L_{\rm IR}$) for our galaxies were derived using three independent indicators: radio, UV and mid-IR.
   We here compare the average of the three indicators for each source
   with the individual derivations for the radio (left panel), mid-IR
   (center panel) and UV (right panel).
   In each plot we show a 0.17~dex error bar, corresponding to the scatter of all individual measurements
   with respect to their average.  $L_{\rm IR}$ is a measure of the ongoing SFR in galaxies: we adopt $L_{\rm
   IR}/L_\odot=SFR/[M_\odot$yr$^{-1}]\times10^{10}$, appropriate for a Chabrier (2003) IMF.
   From left to right in each panel, galaxy IDs are BzK-25536, 4171, 17999, 16000,
   21000, 12591.
           }
              \label{fig:SFR_plots}
    \end{figure*}

Finally, we investigate if the requirement of the availability of a
spectroscopic redshift prior to CO observations could have biased our
sample. Given that most of the spectroscopic redshifts
in GOODS-N were derived from optical spectroscopy, it is possible that
this requirement might have biased our sample against optically faint
galaxies or equivalently, given the near-IR selection of our sample,
to objects with relatively low reddening by
dust. Fig.~\ref{fig:EBV_Mass_CO} shows the amount of reddening for the
BzK galaxies, estimated from the slope of their UV continuum, plotted as
a function of stellar mass. There is a clear trend between $E(B-V)$
and stellar mass in our sample, as reported also by Greggio et
al. (2009). More massive galaxies are more strongly reddened, most
likely a result of the mass-metallicity relation (Erb et al. 2006;
Hayashi et al. 2009; Onodera et al. 2010).  Incidentally, we note that
the vast majority of galaxies with $M\simgt10^{11}M_\odot$ (including
the six CO detected galaxies) are too red to be selected as BM/BX
galaxies from the UV (Fig.~\ref{fig:EBV_Mass_CO}), a result consistent
with what is found by van Dokkum et al. (2006) at slightly higher
redshifts. 
From Fig.~\ref{fig:EBV_Mass_CO} it
is clear that the six CO--detected galaxies lie on the average trend of
$E(B-V)$ versus stellar mass, i.e. they do not show particularly bluer
colors than galaxies of similar stellar mass.

To summarize the characterization of our sample, we conclude that we
have observed a subsample of BzK galaxies with fairly high stellar
masses and toward the lower redshift range of the BzK selection but
with overall physical properties (SFRs, stellar mass and reddening)
that are consistent with those of the general population of near-IR
selected galaxies at $z\sim2$.

   \begin{figure*}[!ht]
   \centering
   \includegraphics[width=13.0cm,angle=0]{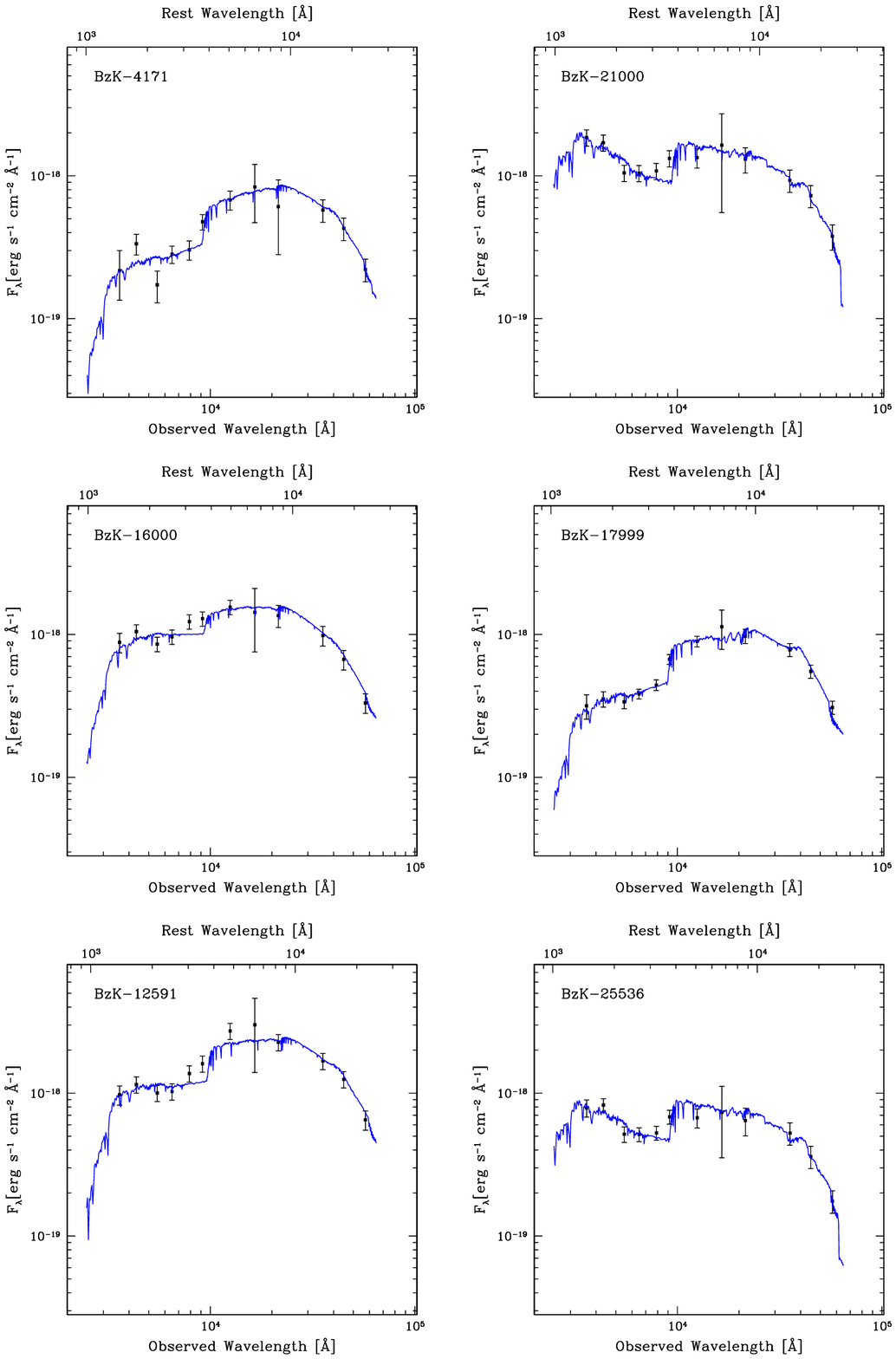}
   \caption{Spectral energy distribution of our target sources showing
   best fitting synthetic templates from the M05 library. We
   consider constant star
   formation rate models with 0.5, 1 and 2 solar metallicities and allow for reddening using a Calzetti et al. (2000)
   law.
        Errors bars in the plots have been increased to formally have $\chi^2_{reduced}=1$.
           }
              \label{fig:sed}
    \end{figure*}

   \begin{figure*}
   \centering
   \includegraphics[width=15.0cm,angle=0]{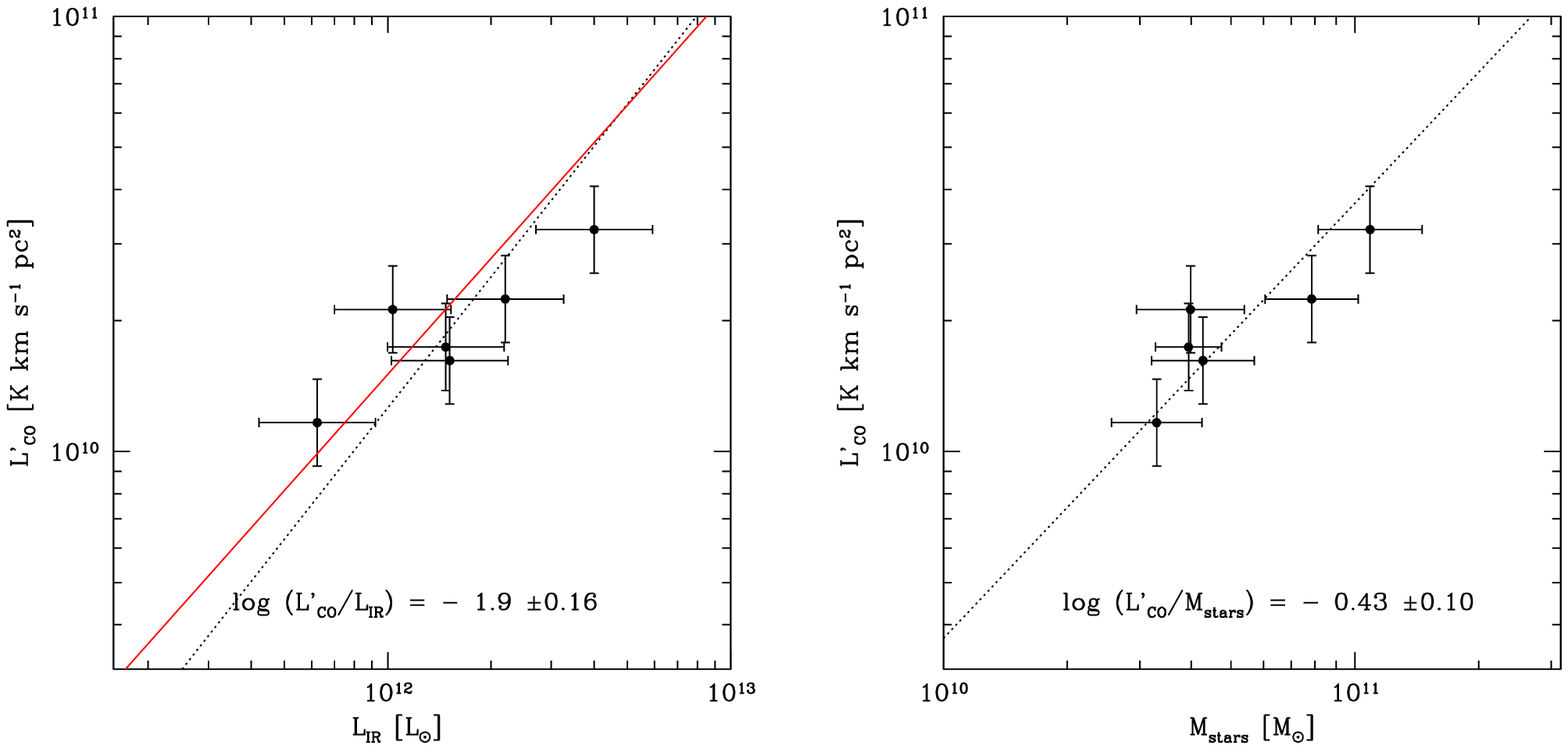}
   \caption{CO luminosities ($L'_{\rm CO}$) versus the bolometric IR
   luminosities ($L_{\rm IR}$, left panel) and versus the stellar
   masses (right panel). The dotted line in each panel is the
   best linear fit. The average ratio and dispersion is
   reported in the bottom of each panel (units are K~km~s$^{-1}$~pc$^2$~$L_\odot^{-1}$ for the ratio in the left 
   panel and K~km~s$^{-1}$~pc$^2$~$M_\odot^{-1}$ in the right panel).
   The red solid line in the left panel correspond to the fit to BzK galaxies at $z=1.5$ and spiral galaxies,
   as shown in the left panel of Fig.~\ref{fig:both_plot_6CO}. In both panels galaxy IDs are, from top to bottom:
   BzK-12591, 21000, 4171, 17999, 16000, 25536.
           }
              \label{fig:xLinear}
    \end{figure*}

\subsection{Total IR luminosities and star formation rates}

In order to accurately derive the SFRs and IR luminosities of our
sample galaxies we have used all the available SFR indicators. This
includes the dust corrected UV luminosities, mid-IR continuum
luminosities from 24$\mu$m Spitzer imaging and the VLA 1.4~GHz radio
continuum. These datasets and the techniques that we use to convert
observed luminosities into SFRs (and equivalently $L_{\rm IR}$; we adopt here
$L_{\rm IR}/L_\odot=SFR/[M_\odot$yr$^{-1}]\times10^{10}$, appropriate for
a Chabrier 2003 IMF) are
described in detail in Daddi et al. (2007) and are largely based on
the Chary \& Elbaz template libraries (but we here use an improved 1.4~GHz
map, Morrison et al. 2010).  A
comparison between these (and other) SFR indicators within the GOODS
sample of BzK selected galaxies for both individually detected sources
and through stacked sources is discussed in detail in Daddi et
al. (2007) and shows that, on average, there is very good agreement between the
different indicators.

We find that the SFRs estimated using the different techniques agree
generally very well also for our CO--detected galaxies, typically to
within much better than a factor of two (Fig.~\ref{fig:SFR_plots}).  In
particular, no galaxy in our sample would be classified as a mid-IR
excess galaxy according to the classification of Daddi et al. (2007b),
which could have suggested the presence of an obscured AGN.  The
largest discrepancy found is for BzK-12591 where we find a radio
estimated IR luminosity that is twice as high as both the UV and
mid-IR estimates. This might hint at a possible presence of an AGN, as
also suggested by the possible [NeV] emission in the optical spectrum.

We decided to adopt the average of the three estimators for the best
measurement of $L_{\rm IR}$ and, subsequently, SFR, with the exception
of BzK-12591 where we used the average of the UV and mid-IR estimates
only (if we had included also the radio its SFR would have increased by
0.17~dex, leaving the remainder of our analysis basically
unchanged). The uncertainty of this measure is derived using the
dispersion of the differences between the average and each individual
determination. From this we find that our SFR should be accurate to
$\sim$0.17~dex.
The SFRs derived from the UV (as used in Fig.~\ref{fig:SFR_Mass_CO}) agree with our best SFRs
within 0.02~dex, with a scatter of 0.14~dex.  

Inspection of the deep 70$\mu$m imaging from the Far Infrared Deep
Extragalactic Survey (FIDEL, Frayer et al. 2009) shows that BzK-21000
is detected with a flux of $4.9\pm0.7\mu$Jy, from the catalog of Magnelli
et al. (2009), as noted also by Casey et al. (2009).
The measured $70\mu$m flux corresponds to 
$L_{\rm IR}=2.1\pm0.3\times10^{12}L_\odot$
using the Chary \& Elbaz (2001) library that we adopt,  in very good
agreement with our estimate in
Tab.~\ref{tab:4}. This is an independent confirmation of our $L_{\rm IR}$
and SFR estimates.

Overall, the galaxies in our CO sample are actively star forming, with
SFRs ranging from about 60$M_\odot$~yr$^{-1}$ for BzK-25536 to
400M$_\odot$~yr$^{-1}$ for BzK-12591 (Tab.~\ref{tab:4}).

\subsection{Stellar masses from synthetic template fitting}

Multicolor photometry in the rest frame UV, optical and near-IR bands
is available for our CO detected galaxies in GOODS-N, as presented by
Daddi et al. (2007). The UBVRIzJHK photometry, plus the photometry in
the first three IRAC bands at 3.6$\mu$m, 4.5$\mu$m and 5.8$\mu$m
is shown for all target galaxies in Fig.~\ref{fig:sed}\footnote{Following Maraston et al. (2006)
and other work we do not use the IRAC 8.0$\mu$m band as it appears to be often contaminated
by dust emission in the BzK galaxies, see also Daddi et al. (2007b).}.

We used synthetic templates from the library of Maraston (2005)
to derive stellar population properties and in particular stellar
masses. These templates have a consistent treatment of the emission of
AGB stars that is important for the correct derivation of stellar
masses in the distant Universe (see, e.g., Maraston et al. 2006).
Given that all of the galaxies in our sample are actively star
forming, we have used templates with a constant SFR, a large range of ages,
Chabrier (2003) IMF and a range of metallicities (half solar, solar
and twice solar, appropriate for the expected range in stellar
masses). We allow for reddening using the Calzetti et al. (2000)
extinction law.  We explicitly account for the mass loss due to
stellar evolution, so that the stellar masses that we derive are lower
by some 10-20\% (depending on age; see Fig.~3 of Maraston et al.
2006) than the time integral of the SFR. The results are summarized in
Tab.~\ref{tab:4}. The stellar masses range from
$3.3\times10^{10}M_\odot$ for BzK-25536 to $1.1\times10^{11}M_\odot$
for BzK-25536.  The fitted stellar masses are only 0.06~dex lower on
average than those from the D04 empirical formula, with a scatter of
0.08~dex. This implies that Fig.~\ref{fig:SFR_Mass_CO} is relevant
also for the stellar masses.

We estimated the
uncertainties in the derivation of the stellar masses on the basis of
$\chi^2$ variations, following Avni (1976) for the case of one
interesting parameter, marginalizing thus on the template
age, reddening and metallicity. We find that, given our accurate
multiwavelength SEDs, the formal (statistics)
 $1\sigma$ error on the stellar masses
are of order of 0.10--0.15~dex. 

\begin{table*}
{\footnotesize
\caption{Physical properties of the CO detected galaxies}             
\label{tab:4}      
\centering                          
\begin{tabular}{l c c c c c c c c c c c c}        
\hline\hline                 
Source & SFR$_{\rm best}^1$ & $L_{\rm IR}$$^2$ & $L'_{\rm CO}$$^3$ & SFE & M$_{\rm stars}$ & M$_{\rm gas}^4$ & $f_{\rm
gas}$$^5$ & M$_{\rm dyn}$$^6$
& $\alpha_{\rm CO}$ & $t_{\rm gas}^7$ & $t_{\rm rot}$~$^8$ & $t_{\rm SFR}$~$^9$ \\
& M$_\odot$~yr$^{-1}$ & $10^{12}L_\odot$ & $10^{10}$K~km~s$^{-1}$~pc$^2$ & $L_\odot$K$^{-1}$km$^{-1}$~s~pc$^{-2}$ &
$10^{10}$M$_\odot$ & $10^{10}$M$_\odot$ &
& $10^{10}$M$_\odot$ & $M_\odot$K$^{-1}$km$^{-1}$~s~pc$^{-2}$ & Gyr & Gyr & Gyr \\
\hline
BzK-4171 & 103 & 1.0 & 2.1 & 49 & 4.0 & 7.7 & 0.66  &  8.2 &  $3.9\pm1.2$ & 0.74 & 0.09 &0.39  \\
BzK-21000 & 220 & 2.2 & 2.2 & 98 & 7.8 & 8.1 & 0.51  & 9.3 &  $2.7\pm1.4$ & 0.37 & 0.17 & 0.36  \\
BzK-16000 & 152 & 1.5 & 1.6 & 93 & 4.3 & 5.9 & 0.58  & -- & -- & 0.35 & 0.34 & 0.28  \\
BzK-17999 & 148 & 1.5 & 1.7 & 85 & 3.9 & 6.3 & 0.62  & 7.3 &  $4.0\pm1.3$ & 0.43 & 0.13 & 0.27  \\
BzK-12591 & 400 & 4.0 & 3.2 & 124 & 11 & 12 & 0.52  & --  &   -- & 0.29 & 0.09 & 0.27 \\
BzK-25536 & 62 & 0.6 & 1.2 & 53 & 3.3 & 4.2 & 0.56  & -- & -- & 0.68 & 0.21 & 0.53  \\
\hline                                   
\end{tabular}\\
Notes:\\
1: SFR$_{\rm best}$ is the average of dust corrected UV, mid-IR and radio derived SFRs. Typical errors are at the
level of 0.17~dex (see text).\\
2: we use $L_{\rm IR}/L_\odot=SFR/[M_\odot$yr$^{-1}]\times10^{10}$, adapted from Kennicutt et al. (1998) using a
Chabrier IMF.\\
3: we apply a correction of 16\% to account for the CO[2-1] transition being slightly sub-thermally excited;
Dannerbauer et al. (2009).\\
4: computed from $L'_{\rm CO}$, assuming $\alpha_{\rm CO}=3.6$ as deduced from our observations\\
5: $f_{\rm gas} = M_{\rm gas}/(M_{\rm gas}+M_{\rm stars})$; the average error is 9\% (including also the uncertainty
on $\alpha_{\rm CO}$\\
6: estimated for galaxies resolved in CO emission, except BzK-16000 that is seen nearly face-on;
errors on $M_{\rm dyn}$ range from 30 to 50\%.\\
7: $t_{\rm gas} = M_{\rm gas}$/SFR\\
8: $t_{\rm rot} = 2\pi r_e/(v_{\rm FWHM}/2)$ is the rotation time at the half light radius\\
9: $t_{\rm SFR} = M_{\rm star}$/SFR\\
}
\end{table*}

\section{Star formation efficiencies}
\label{sec:LCO}

From the observed CO[2-1] fluxes we have derived CO luminosities
$L'_{\rm CO}$ spanning a range of
1.2--3.2$\times10^{10}$~K~km~s$^{-1}$~pc$^2$ (Tab.~\ref{tab:4}; see
Solomon \& van den Bout 2005 for the conversion formula in the adopted
units). We here have applied a 16\% correction to convert luminosities
derived from CO[2-1] to $L'_{\rm CO}$ (defined relative to the
fundamental CO[1-0] transition) based on the BzK excitation results in
Dannerbauer et al. (2009). Given the flux measurement errors and the
uncertainty in this conversion, we estimate typical errors in $L'_{\rm
CO}$ to be at the level of 0.1~dex.

We compare the CO luminosities to the $IR$ luminosities and stellar
masses of our sample galaxies in Fig.~\ref{fig:xLinear}. We find that
the CO emission
correlates well with both $L_{\rm IR}$ (hence SFR) and stellar
mass (we exclude at the 2.5$\sigma$ [3$\sigma$] confidence level that the CO emission does
not correlate with SFR [mass]).  The ratio of $L'_{\rm CO}/L_{\rm
IR}$ (i.e., SFE) has a scatter of only 0.16~dex, while the individual values
are all within a factor of 2.5. The scatter is in
fact similar to the error of $L_{\rm IR}$ only, which possibly
dominates the total error (this would suggest even lower intrinsic
scatter in $L_{\rm IR}$/ $L'_{\rm CO}$).  The ratio of $L'_{\rm
CO}/M_{\rm stars}$ has an even lower scatter of 0.10~dex. This is
driven by the fact that the formal errors on stellar masses are lower
than those on IR luminosities. This correlation of the CO luminosity with {\em both}
the SFR and stellar mass is not surprising, given that the six galaxies lie on top
of the stellar mass-SFR correlation and their SSFR has a scatter of only 0.12~dex.
However, it is remarkable that we can now probe for the first time that the narrow range
of SSFR corresponds to a similarly narrow range of SFEs, i.e. is directly related to the gas
properties.

   \begin{figure*}
   \centering
   \includegraphics[width=15.0cm,angle=0]{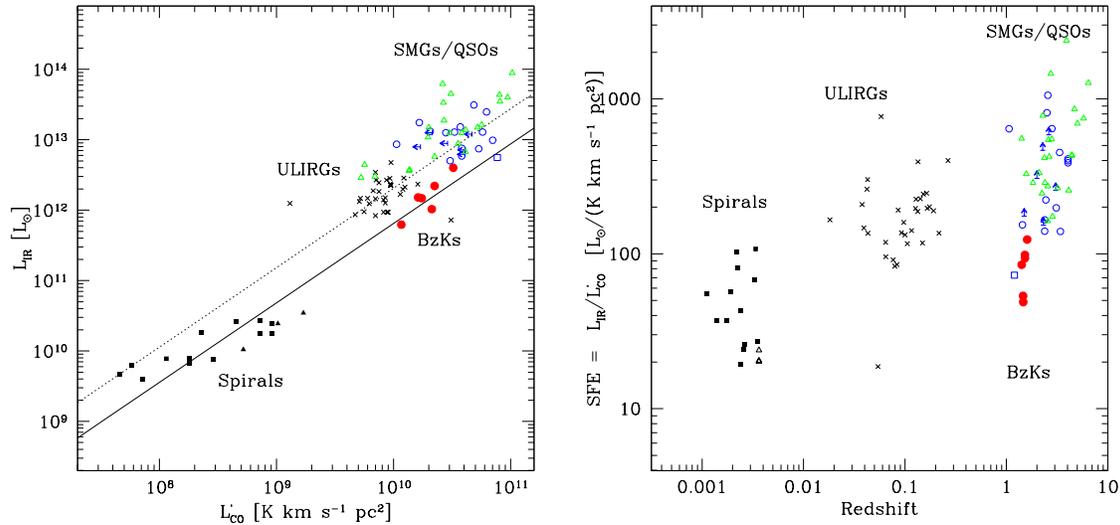}
   \caption{Comparison of the CO and IR luminosities of the CO--detected
   BzK galaxies (red filled circles) to various local and distant
   samples: SMGs (Greve et al. 2005 and Daddi et
   al. 2009: blue empty circles; Frayer et al. 2008: blue empty square),
   QSOs (Riechers et al. 2006; Solomon \& van den Bout 2005; green empty triangles), local ULIRGs (Solomon et al. 1997:
   crosses), local spirals (Leroy et al. 2008; 2009: filled
   squares;
   Wilson et al. 2009: filled triangles). For local spirals we associate a Hubble-flow redshift (right panel)
   based on the actual distance.
   The solid line in the left panel shows the best fitting $L'_{\rm CO}$ versus $L_{\rm IR}$ relation to the
   combined sample of local spirals and distant BzK galaxies. The dotted line shows the same relation with
   normalization higher by  0.5~dex.
           }
              \label{fig:both_plot_6CO}
    \end{figure*}

In Fig.~\ref{fig:both_plot_6CO} we compare the CO and IR luminosities
of the six CO--detected BzK galaxies with other cosmologically relevant
galaxy populations: SMGs (Greve et al. 2005; to which we add more
recent results of Daddi et al. 2009ab and Frayer et al. 2008),
quasars (QSOs; Riechers et al. 2006 Solomon \& van den Bout 2005),
 local
ULIRGs (Solomon et al. 1997) and local spirals taken from the HERACLES
survey (Leroy et al. 2008; 2009) and from the Virgo Cluster survey of
Wilson et al. (2009). 
We notice that in Daddi et al. (2008) we used
the results of Yao et al. (2003) to define a bona fide comparison
sample of local spirals, following Solomon \& van den Bout (2005) and
Greve et al. (2005).  While the results and conclusions of this paper
would remain the same if we were still using the Yao et al. (2003)
sample, we here prefer the more recent samples as the Yao et al. sample
was selected in the submm.
Also, we here chose not to adopt the IR luminosities for SMGs
estimated by Greve et al. (2005) but use a lower conversion factor
between $850\mu$m flux and IR luminosity to account for the evidence
that their SEDs are colder than previously thought (e.g., Pope et
al. 2006; 2008; Kovacs et al. 2006; Daddi et al. 2009). In addition, we exclude lower
luminosity objects from the Solomon et al. (1997) sample requiring
$L_{\rm IR}>0.7\times10^{12}$ to obtain a sample of 
ultra-luminous IR galaxies with IR luminosities similar to the BzK galaxies.

Based on our sample of six galaxies, we thus confirm and extend the
result of Daddi et al. (2008) 
that the typical SFE of massive BzK galaxies is
similar to that of spirals and significantly lower than that of
SMGs.  We find that the average SFE (in units of
$L_\odot$K$^{-1}$~km$^{-1}$~s~pc$^{-2}$; the errors given in the following 
are the uncertainties of the mean) of BzK galaxies is $84\pm12$, that
of SMGs is $390\pm60$, for QSOs we find $590\pm100$, for local ULIRGs we have $250\pm30$ and local
spirals have $48\pm7$.  Therefore, the typical BzK galaxy has about
80\% higher SFE of that of a local spiral, but individual values are well within
the range measured for local spirals even though BzKs are a 
hundred times more luminous.  
Local ULIRGs have SFEs that are
three times larger on average than BzK galaxies. Although 
there is some overlap in the spread of individual
targets,  one can see that  the two distributions do not really overlap in the
$L_{\rm IR}$ versus $L'_{\rm CO}$ plane (Fig.~\ref{fig:both_plot_6CO}).

For SMGs/QSOs the average SFE is 4.5/7 times larger than that of an average
BzK galaxy. Although at least one case is known of an SMG with a SFE
at the level of that of BzK galaxies (Frayer et al. 2008), for the
vast majority of the sample the distribution of values are fairly
disjoint. We note that, in contrary to our BzK galaxies, the CO
luminosities of SMGs are derived from high-$J$ transitions, and could
be underestimated if a substantial amount of cold gas is present.  Such
a correction could indeed be important in some cases (Greve et al.
2003; Carilli et al. 2010) but the difference still
appears to be quite significant if a correction was applied using the
average CO SEDs of SMGs (Weiss et al. 2005; 2007) or local ULIRGs
(Papadopoulos et al. 2007). 

Given the comparable SFE of local spirals and BzK galaxies, and the
evidence that indeed the BzK galaxies are gas rich disks in the
distant Universe, we can formally consider them as a homogeneous
population and derive the best fitting $L'_{\rm CO}$ versus $L_{\rm
IR}$ relation for the combined population. We find the relation:

\beq
{\rm log} (L_{\rm IR}) = 1.13\times({\rm log} L'_{\rm CO}) + 0.53
\eeq

\noindent
with $L'_{\rm CO}$ in units of K~km~s$^{-1}$~pc$^2$ and $L_{\rm IR}$
in units of $L_\odot$.  Both parameters in the fit are determined with
an accuracy of 0.07. The relation appears to fit both the sample of
BzK galaxies (see also Fig.~\ref{fig:xLinear}) and local spirals well,
with a root mean square scatter of 0.24~dex.  
The local ULIRGs and SMGs are systematically offset from the fit,
but are consistent with a relation having a similar slope but with
a factor of 3--4 higher normalization.

Similarly, Fig.~\ref{fig:both_plot_6CO} shows that our BzK
galaxies should have been factors of 3--4 fainter in their CO fluxes
had they followed  the trends previously established for 
equivalently luminous galaxies such as local ULIRGs and
SMGs, as discussed in the introduction. Had that been the case, it would
have taken $\simgt10$ times longer integrations to detect these
sources, which would have made detections impossible until ALMA is in
full operation.

\section{Gas kinematics in clumpy disks: numerical simulations}
\label{sec:models}

   \begin{figure}[ht]
   \centering
   \includegraphics[width=8.8cm,angle=0]{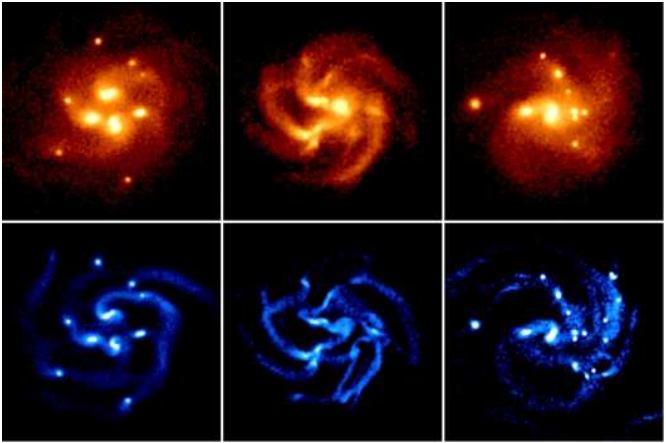}
   \caption{
	   A few examples of our simulations of clumpy disk galaxies, based on the
	      results of Bournaud, Elmegreen \& Elmegreen (2007). 
	      Top panels show the distribution of stellar light as observed in the UV
	      (tracing mainly star formation),
	         bottom panels that of the molecular gas. The distribution of star formation and optical morphologies
		 are dominated by giant clumps, and our models include a variety of clumpiness and bulge fractions
		 consistent with observed star-forming disks at high redshift. 
		 These snapshot show a $40 \times 40$~kpc area ($5''\times5''$ at $z=1.5$).
           }
              \label{fig:FB_models}
    \end{figure}

A critical piece of information that our observations can provide is the derivation of the dynamical mass of the systems
through the observed gas kinematics and spatial extent. In today's disk galaxies, the cold gas component is slowly
rotating and directly traces the dynamical mass, which relates to the circular velocity squared and the radius of the
system (e.g., de Blok et al. 2008). However, gas motions in high-redshift disk galaxies are more complex. 
In particular, larger
 molecular gas reservoirs are present and optical morphologies are
dominated by giant star-forming clumps. This is seen in our sample (Fig.~\ref{fig:3color}) as well as in the majority
of star-forming disks at high redshift (Elmegreen et al. 2007). These giant clumps are up to $\sim 1000 \times$ more
massive than the star-forming clouds of today's spirals (Elmegreen et al. 2009b). Such gas-rich clumpy disks can have
significant 
non-circular motions (Immeli et al. 2004; Bournaud et al. 2008) and show strong turbulence in their ionized
gas (e.g., Genzel et al. 2008; F\"orster-Schreiber et al. 2009). It is likely that 
such motions are also present in their cold gas (CO)
component (Bournaud \& Elmegreen 2009; Dekel et al. 2009b). Clumpiness and turbulence influence the kinematics,
sizes, velocity fields, and dynamical mass estimates of distant galaxies (Bournaud et al. 2008; Burkert et al. 2009).

To recover dynamical masses accurately from CO velocities and to estimate associated uncertainties, we used numerical
simulations of gas-rich clumpy galaxies based on the work of (Bournaud, Elmegreen \& Elmegreen 2007; hereafter
BEE07). These simulations start as gas-rich, gravitationally unstable disks that rapidly evolve into
clumpy disks with a phase space distribution similar to cosmological simulations of clumpy disk formation (Agertz et
al. 2009; Ceverino et al. 2009) and consistent with observed clumpy galaxies (see BEE07 and Bournaud et al. 2008). The initial conditions ($t=0$) of the six new models used here were scaled to match the typical masses and sizes of the
observed galaxy sample, and are as follows: 
stellar masses of 6$\times10^{10}$ M$_\odot$, initial half mass radii of 5~kpc, gas fractions $f_{\rm g} = (M_{\rm gas}/M_{\rm gas}+M_{\rm stars})$ of 30, 50, and 70\%, and dark matter fractions
$f_{\rm DM} = (M_{\rm dark}/(M_{\rm dark}+M_{\rm gas}+M_{\rm stars})$ of 15 and 35\% within the effective radius.  The output of the simulations were retained after 250, 450, 650Myr, which, in the evolutionary sequence studied in BEE07, covers all stages from very clumpy bulgeless morphologies to smoother bulgy systems, consistent with our observed sample (Fig.~\ref{fig:FB_models}; compare to Fig.~\ref{fig:3color}). In each case the system was {\em observed} with $30^o$ and $60^o$ inclinations and at three different azimuths, giving a total of 108 ``independent'' projections.

   \begin{figure}
   \centering
   \includegraphics[width=8.8cm,angle=0]{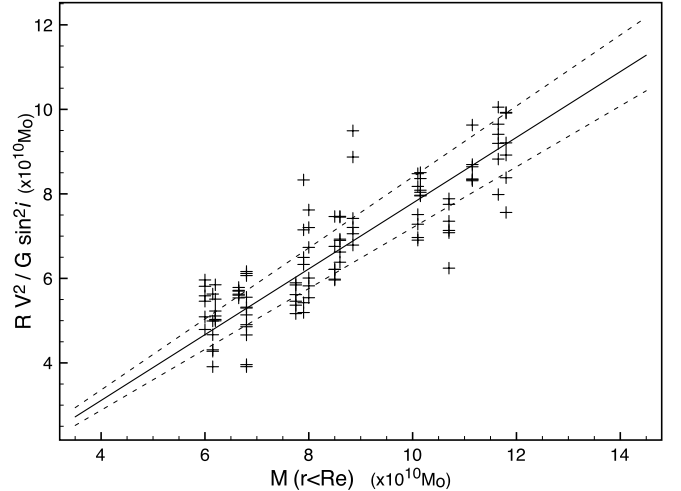}
   \caption{
	   Calibration of the relation between the $R\times V^2$ estimator corrected for inclination effects (y-axis) and
	         the effective amount of mass contained within the half light
		    radius ($r_e$, x-axis), based on clumpy disk models. Sequences of points at fixed dynamical mass
		       correspond to the same model observed with different inclinations and azimuths.
		           The solid and dashed lines show the best fitting relation (Eq.~\ref{eq:Mdyn})
			       and the $1\sigma$ scatter on the relation, respectively.
           }
              \label{fig:FB_Mtrue_Mobs}
    \end{figure}

Taking advantage of the fact that the actual mass encompassed at the effective radius $r_e$ can be directly measured
in the simulations, we compare its value
to a crude estimate of the dynamical mass, $(v_{\rm FWHM}/2)^2 \times r_e/ (G\, \sin^2
i)$, in order to derive a robust method to estimate dynamical masses from high-redshift CO spectra. For each
projection of the simulations, we derive the spatial distribution of ``molecular'' gas component. To do so, we only
consider 
gas in regions where the density is higher than 100 atoms\,cm$^{-3}$~\footnote{This selection actually includes
almost all the gas mass in the models, since the gas fraction is
high and the density is high almost everywhere.}. We measure the effective radius $r_e$ not as the true half-light
radius but as the effective radius of a S\'ersic fit to the radial profile distribution, just as in our observations. We
derive synthetic CO spectra from which we estimate the FWHM velocity $v_{\rm FWHM}$. We also produce UV stellar images
from the simulations to reproduce deprojection uncertainties and compare UV and CO sizes (see below). 
In this way, the $(v_{\rm FWHM}/2)^2 \times r_e/ (G\, \sin^2 i)$ ratio is computed with the same definitions,
assumptions and possibly biases as in the observations. The main result of this exercise is shown in Fig.~\ref{fig:FB_Mtrue_Mobs}. We find the following best fitting relation to the actual mass inside $r_e$:

\beq
M(r<r_e) = 1.3 \times \frac { r_e \times (v_{\rm FWHM}/2)^2 } {G \ \ {\rm sin}^2\ i} \;\;\;\;\;\; \pm 12.5 \%
\label{eq:Mdyn}
\eeq

The {\it correction factor} of 1.3 includes various effects:

\begin{itemize}
\item the flattening of the baryonic mass distribution in a disk (Binney \& Tremaine 2008). This has a small effect
(around $-5\%$) on the total ($+30 \%$) correction factor. 

   \begin{figure}[!ht]
   \centering
   \includegraphics[width=8.8cm,angle=0]{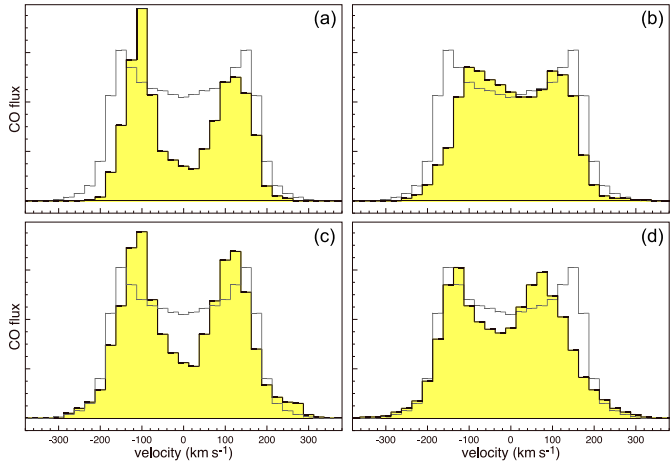}
   \caption{
Four CO spectra extracted from our simulations. The grey histogram shows the initial ($t=0$) uniform disk, the
black lines and yellow histograms correspond to evolved stages in the simulations (i.e., realistic clumpy disks).
Because of the discrete sampling of the velocity field related to the clumpiness of the gas, and the individual
motions of each clump, the predicted spectra are more irregular than for $z=0$ spiral galaxies. Overall, the
typical double-horn profile is preserved, but there can be a strong central dip at zero velocity (a,c, resembling the
observed spectrum of BzK-4171 in Fig.~\ref{fig:1D}) or, on the other hand, relatively flat spectra with little zero
velocity central depression
(b, as in BzK-17999). Asymmetrical spectra (a,d) are also found, as observed in BzK-12591 and BzK-25536.
The velocity width of these spectra is smaller than for the initial smooth rotating disk. This corresponds to clump
evolution and migration with rotation velocities below the circular velocity, and the associated turbulent pressure
support that compensates for the lack of rotation. This lies at the origin of the calibration factor in
Eq.~\ref{eq:Mdyn}.
           }
              \label{fig:FB_spectra}
    \end{figure}
\item a possible bias on estimating the projection angle $i$ because of the clumpiness of these galaxies. We
de-projected simulated UV images just like observations, so that any uncertainty and bias on $i$ are included in
our results (Eq.~\ref{eq:Mdyn} and the scatter of Fig.~\ref{fig:FB_Mtrue_Mobs}). We actually found that the
inclination $i$ retrieved from the UV images tends to 
slightly 
overestimate\footnote{For instance, a face-on disk ($i = 0$) 
could have a large fraction of its light distributed
in a few main clumps, leading to an axis ratio $b/a < 1$ and hence a derived inclination $i > 0$.}
the actual projection angle (which is known in simulations). This is also a small effect, around $+10\%$ on $M(r<r_e)$.

\item The fact that gas motions largely differ from circular orbits is the main effect, around $+25\%$ on $M(r<r_e)$.
The giant and massive gas clumps interact with each other, undergo dynamical friction and migrate radially (BEE07,
Dekel et al. 2009b). 
The presence of a high velocity dispersion $\sigma$ decreases the observed rotation velocity $v_{\rm rot}$ compared to the actual circular velocity 
$v_{\rm circ}$ tracing the potential, 
with $v_{\rm circ}^2 \simeq v_{\rm rot}^2 + 3\sigma^2$, where $\sigma$ is the one-dimensional gas velocity dispersion (observed line-of-sight dispersion)
and the factor 3 is for isotropic gas turbulence in three dimensions.
The effect is also known as `asymmetric
drift' in stellar kinematics, and the equivalent of a turbulent pressure support in addition to the rotational
support for a gas disk (Burkert et al. 2009; Elmegreen \& Burkert 2009). The velocity dispersion $\sigma$ of the CO component cannot be directly
measured given the current observations, but should be several tens of km~s$^{-1}$ based on H$\alpha$ data (e.g., F\"orster-Schreiber et al. 2009)
and observed clump sizes (which is set by the Jeans mass, Bournaud \& Elmegreen 2009). The theoretical correction
factor for the mass expected from a disk with a circular velocity of about
 200~km~s$^{-1}$ is then consistent with the $+25\%$ correction from our simulations. 
\end{itemize}

Our simulations also have the advantage of including
 the clumpiness of the gas resulting from an irregular sampling of
the underlying velocity field and a dispersion in the dynamical mass measurements -- a given model has different
velocity widths $v_{\rm FWHM}$ when we observe it from different azimuths. The resulting uncertainty on the estimated
mass is nevertheless relatively small, about 12.5\% (this uncertainty accounts for the method used and clumpiness of
disks). In summary, 
the presence of giant clumps does not significantly
 hamper the possibility to estimate accurate dynamical masses.

The models were started with similar gas and stellar sizes, and we checked that the half light radii derived from the
UV stellar emission and from the CO emission remain similar during the evolution (Fig.~\ref{fig:FB_models}). This is
in agreement with what is found for our sources (see Section~\ref{sec:data}) as well as local disks (Leroy et al. 2008). This implies that, referring to the dynamical mass within the half light radius, the quantity computed according to Eq.~\ref{eq:Mdyn} should be compared to {\em half} of the total amount of stellar mass and gas mass in the system, plus the amount of dark matter eventually present within $r_e$, i.e.:

\beq
M(r<r_e) = 0.5 \times \left ( M_{\rm star} + M_{\rm gas} \right ) + M_{\rm dark} (r<r_e)
\label{eq:Mdyn2}
\eeq

Applying this result to observations, we derive $M(r<r_e)$ from Eq.~\ref{eq:Mdyn}, subtract the known stellar mass $M_{\rm star,tot}$ and an estimate of $M_{\rm dark} (r<r_e)$ to finally derive the gas mass $M_{\rm gas,tot}$ and compare to the observed CO fluxes.

Note that we calibrated the measurements (Eq.~\ref{eq:Mdyn} and Fig.~\ref{fig:FB_Mtrue_Mobs}) at a radius $r=r_e$ and using the $v_{\rm FWHM}$ velocity, in order to match observational measurements. Dynamical masses estimates at different radii should be similar, as the main contribution to these factors in Eq.~\ref{eq:Mdyn}, namely clumpiness and high turbulence, are present all over high-redshift disks.

The numerical simulations spanned gas fractions from 30 to 70\%. Reproducing clumpy structures similar to those
observed (Fig.~\ref{fig:3color} ~and~\ref{fig:acs}) requires high gas fractions and disk masses (Bournaud \&
Elmegreen 2009), and we will show that our observations support a high molecular gas fraction in these galaxies. In
any case, kpc-sized clumps are observed, which imply a large Jeans length and a high turbulent speed (velocity
dispersion). As the underlying physics in Equation~\ref{eq:Mdyn} is based on the
gas velocity dispersion, it should remain valid independent of the actual gas fraction.

\subsection{Explaining the shapes of the observed CO spectra}

An application of the numerical models is to produce synthetic CO spectra and compare their shapes to the observed ones. 
Indeed, the observed spectra have some particularities that would not be expected for regularly
rotating disks such as seen in $z=0$ spirals. 

   \begin{figure*}
   \centering
   \includegraphics[width=15cm,angle=0]{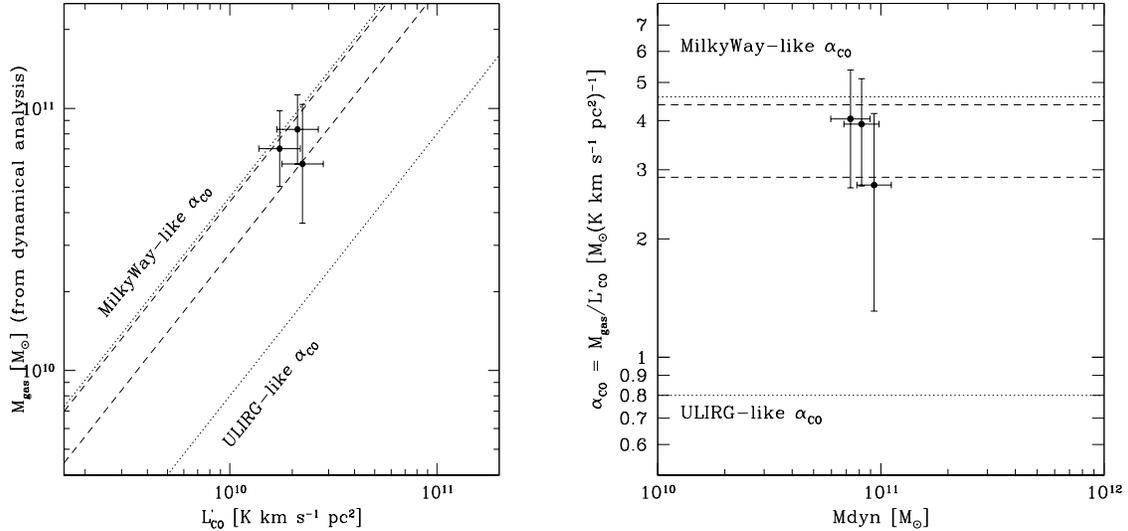}
   \caption{Constraints on the CO luminosity to gas mass conversion factor ($\alpha_{\rm CO}$).
   The analysis was possible for three galaxies for which we plot gas masses
   versus CO luminosities (left panel) and their ratio 
($\alpha_{\rm CO}$) versus the dynamical masses (right panel).
   The dashed lines show the $1\sigma$ range on  $\alpha_{\rm CO}$ derived by combining the measurements
   for the three galaxies. The dotted lines  show the conversion factor for the Milky~Way galaxy (top) and for
   local ULIRGs (bottom). From left to right galaxies are: BzK-17999, 4171, 21000.
           }
              \label{fig:Mdyn}
    \end{figure*}

For example, the CO spectrum of BzK-4171 (Fig.~\ref{fig:1D}) has a very marked and quite symmetric double horn
profile with a pronounced central depression in the center, falling to only 15\% of the peak flux. Gas spectra in
local galaxies show double horn profiles but the emission at zero velocity never falls below 50\% of the peaks (e.g.,
Walter et al. 2008). The observed feature in BzK-4171 can not be easily explained by assuming that the molecular gas is in a ring: removing molecular gas from the system barycenter would still leave substantial amount of emission at zero velocity, coming from regions where the gas rotation velocity is perpendicular to the line of sight. For instance, even a pure thin ring rotating with $V/\sigma=10$ would have a zero velocity emission at 42\% of the peaks.
Fig.~\ref{fig:1D} shows  that also BzK-12591 and BzK-25536 have significantly asymmetric spectra, with a flux in one
horn much higher than in the other horn. The spectrum of BzK-17999 is almost flat around the central velocity.

The synthetic CO spectra from our numerical simulations shed light on this situation. Spectra qualitatively similar
to BzK-4171 are found, as well as spectra where the zero velocity level is quite high, and double horn profiles that
are strongly asymmetric (Fig.~\ref{fig:FB_spectra}). The explanation is that much of the CO emission comes from a few
big clumps, and therefore is  {\em discretized}. At particular projections, there can be a lack (or an excess) of zero-velocity emission, when few (many) clumps have velocities perpendicular to the line of sight. Significant asymmetries arise when the clumps are not uniformly distributed between the blueshifted and redshifted parts of the disk.

The successful reproduction of the observed spectra give confidence that our
models are correct, at least to first order.
More generally, both observations and modeling show
that turbulent and clumpy disks at high redshift can have integrated spectra that are
distinct from what is seen in the local Universe.

An interesting characteristic in the simulated spectra is that, in spite of their
disturbed shapes, the velocity spread remains about the same in the two horns
(red-- and blue--shifted sides). One of the two sides can have a much higher flux, but not a much larger velocity spread. 
This is seen also in the observed spectra.

\section{Estimating gas masses in normal high redshift massive galaxies}
\label{sec:Mdyn}

Based on the results shown in Fig.~\ref{fig:FB_Mtrue_Mobs} and using
Eqs.~\ref{eq:Mdyn} and \ref{eq:Mdyn2}, we proceed to estimate the dynamical mass within $r_e$ of the CO--detected BzK galaxies, based on the $v_{\rm FWHM}$ velocity from CO[2-1] spectra, the stellar half light radius (that corresponds to
 the CO half light radius), and the inclination $i$ derived from the $b/a$ axis ratio seen in the ACS imaging. Values and uncertainties are given in Tab.~\ref{tab:2}~and~\ref{tab:3}. We limit our analysis to the  CO-detected galaxies that were resolved in CO. We did not estimate a dynamical mass for BzK-16000 because 
the correction due to its inclination is quite uncertain ($b/a \simeq 1$).

In order to estimate the gas mass following Eq.~\ref{eq:Mdyn2}, we account for a dark matter fraction within $r_e$ of
25\% (see also next section). 
Disk galaxies are known to be dominated, at their effective radius, by baryons more than by dark matter (Bosma
1981; de~Blok et al. 2008; Trachternach et al. 2008 -- dark matter dominates rotation curves only at much larger
radii). This implies that the dark matter fraction is {\em well below} 50\%, and there is no expectation for a major
change with redshift (see also F\"orster-Schreiber et al. 2009). The structural properties of $z \sim 2$ galaxies
also indicate that, within their visible, star-forming extent, they have baryonic masses and disk masses that should
account for at least two thirds of the total masses (e.g., Bournaud \& Elmegreen 2009).

   \begin{figure*}
   \centering
   \includegraphics[width=15.0cm,angle=0]{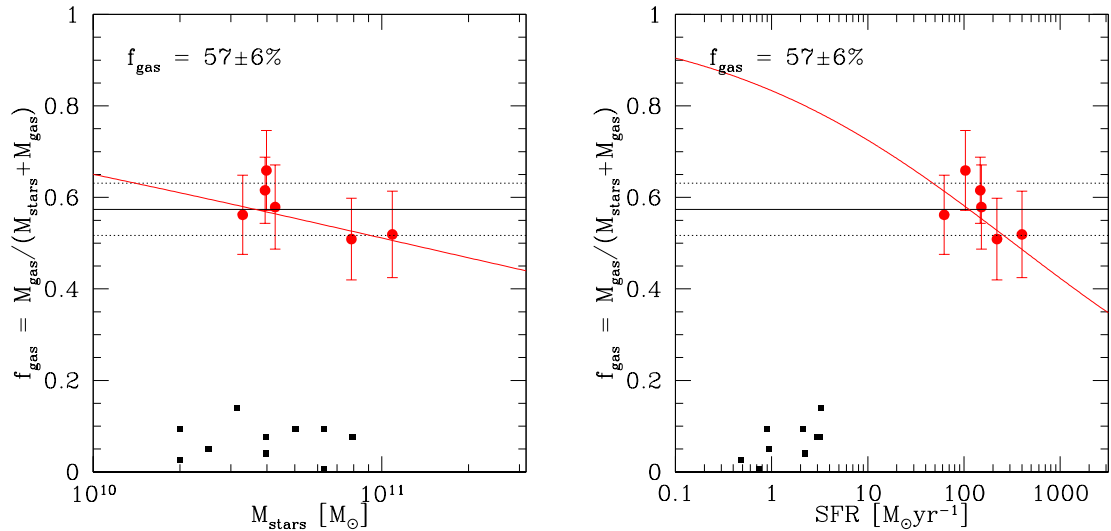}
   \caption{Gas fractions for the six CO detected BzK galaxies, derived
   using a MW-like $\alpha_{CO}=3.6$ are compared to stellar masses (left
   panel) and SFRs (right panel). The solid and dotted horizontal
   lines show the average value and the $1\sigma$ scatter. The red
   curves show the expected trend assuming at face value
   the correlation between $L'_{\rm CO}$ and $L_{\rm IR}$ (hence SFR)
   and the stellar mass-SFR correlation at $z\simgt1.4$ (see discussion in the text). 
   The black filled squares in the left panel show the molecular gas fraction for
   local spirals with $M>2\times10^{10}M_\odot$ from Leroy et al. (2008),
   including H$_2$  and helium (for our BzK galaxies we expect that most of
   the hydrogen is molecular).
   From bottom to top the galaxy IDs are: BzK-21000, 12591, 25536, 16000, 17999, 4171.
           }
              \label{fig:xFrac}
    \end{figure*}

We find dynamical masses within $r_e$ for the three galaxies in the range of 7--9$\times10^{10}M_\odot$ with typical
uncertainties at the level of 30\%. Subtracting the stellar and dark mass from the dynamical mass (Eq.~\ref{eq:Mdyn2}), we derive a total gas mass of order 6.1--8.3$\times10^{10}M_\odot$ for the three galaxies. After propagating all the estimated uncertainties in our measurements and adding in quadrature the 12.5\% uncertainty associated to Eq.~\ref{eq:Mdyn}, the uncertainty in the total gas mass estimate is about 30\% for BzK-4171 and 17999 and 50\% for BzK-21000. 

These gas mass estimates are derived independently from the observed CO luminosities. Thus taking the ratio of the
gas mass estimates and the CO luminosities, we can derive a conversion factor $\alpha_{\rm CO} = M_{\rm gas}/L'_{\rm CO}$ for each system (see Tab.~\ref{tab:4} and Fig.~\ref{fig:Mdyn}), and find a combined result of
\beq
 \alpha_{\rm CO} \ = \ 3.6 \ \ \pm \ 0.8 \ \ M_\odot ({\rm K\ km\ s}^{-1} {\rm pc}^2)^{-1}
\eeq
Although the uncertainty in the individual
$\alpha_{\rm CO}$ estimates is significant, their $\alpha_{\rm CO}$ value are well
above the 0.8~$M_\odot$~(K~km~s$^{-1}$~pc$^2$)$^{-1}$ factor that is generally applied to ULIRGs and SMGs (Downes \&
Solomon 1998; Solomon \& van den Bout 2005; Tacconi et al. 2006; 2008) with individual values for ULIRGs typically in
the range of 0.3--1.3. The individual values are also somewhat lower than the MW factor of 4.6~$M_\odot$~(K~km~s$^{-1}$~pc$^2$)$^{-1}$, although still consistent within $\sim1\sigma$ for each object. 

Combining the three galaxies allows us to exclude the ULIRG-like conversion factor at the 3.7$\sigma$ level, while the MW-like value is within 1.3$\sigma$ of our estimate. 

This conversion factor in turn allows us to estimate the gas content in the BzK galaxies for which a direct estimate was unfeasible due to larger
 observational uncertainties. The gas masses and fractions in these three galaxies turn out to be similarly high as in the three galaxies with dynamical estimates (Tab.~\ref{tab:4}).

We notice that our $\alpha_{\rm CO}$ traces the total amount of gas,
including molecular and atomic hydrogen as well as helium in their respective
proportions within the galaxies half light radii. However, we
expect the vast majority of the hydrogen to be molecular
given that the high observed densities and expected pressures in the interstellar medium of these galaxies (see,
e.g., Blitz \& Rosolowsky 2006; Leroy et al. 2008; Obreschkow \& Rawlings 2009). This is also consistent with hydrodynamic simulations where more than $90\%$ of the gas mass is molecular, at densities above 100~cm$^{-3}$ and temperatures below $\sim 100$~K (Delaye, Bournaud et al., in preparation).

In summary, our analysis shows the existence of large molecular gas
masses in our targets, and a high CO--to--gas mass conversion factor, similar to the one
found in the Milky~Way.  We can estimate the gas mass fractions $f_{\rm gas} = M_{\rm gas}/(M_{\rm gas}+M_{stars})$. 
For consistency, we apply the average
conversion factor of $\alpha_{\rm CO} = 3.6$ to all observed CO
luminosities. The gas fractions range from 51 to 66\%, with the average being 57\% with a dispersion within the sample of of 6\%. 
These galaxies clearly have very high gas fractions, with gas accounting for 
masses that are similar or greater than those of the stars. 

The gas fraction appears quite constant within the sample 
(Fig.~\ref{fig:xFrac}). 
Given the correlation between $L'_{\rm CO}$ (and thus gas mass) and $L_{\rm IR}$ (or SFR) with a slope above unity,
and the correlation between stellar mass and SFR with a slope slightly below unity, one could expect a gas mass
fraction slowly decreasing with increasing stellar mass and SFR. The data are consistent with such a weak trend. 

For comparison, the total gas fractions (including HI, H$_2$ and helium) in local spiral galaxies with the same mass 
is about 20\% (Leroy et al. 2008), and about 7\% if including only H$_2$  and helium (i.e., restricting to the stellar disk). We thus observe a very strong
evolution in the gas fractions of disks at fixed stellar mass. Assuming that most of the gas within $r_e$ in BzK galaxies is molecular, this corresponds
to an average increase of molecular gas content of a factor of 18, at fixed stellar mass, from $z=0$ to 1.5. This is very similar to the  increase
of the SSFR (or the normalization of the stellar mass-SFR relation) between $z=0$ and 1.5. Hence the increase in SSFR can be consistently explained
by the increase of molecular gas content inside star forming galaxies.

Dividing the gas masses by the SFRs we estimate gas consumption timescales  in the range of 0.3--0.8~Gyr. 
The SFRs can in principle continue for even 
 longer timescales, once accounting for gas accretion. This confirms that
star formation can go on for much longer timescales than expected in rapid, generally merger driven bursts. 
The (rotation) dynamical timescale at the half light radius is $t_{\rm rot} = 2\pi r_e/(v_{\rm FWHM}/2)$, and ranges over 0.1--0.3~Gyr
for our galaxies.  On average the gas consumption time thus corresponds to 3.5 rotation timescales for the disks at $r_e$.

\subsection{Discussion of systematic uncertainties}

It is important to investigate possible systematic effects that could
be biasing our dynamical mass analysis and the resulting derivation of the
$\alpha_{\rm CO}$ conversion factor. Generally speaking, these CO
detected BzK galaxies have bolometric IR luminosities similar to those
of local ULIRGs, but we advocate a much larger conversion
factor. Several reasons make this quite plausible as these galaxies behave like disk galaxies in many regards --
see discussions in Daddi et al. (2008), Dannerbauer et
al. (2009) and in the next section. However, we should investigate if
there are issues that could have led to an overestimation of the
conversion factor.

A first possible issue deals with systematics in the derivation of the
dynamical masses. Our analysis is calibrated on clumpy disk numerical
modeling, which seems well justified given that our sources are
observed to be highly clumpy. 
If we were to neglect the factor of 1.3 in Eq.~\ref{eq:Mdyn}, the
combined constraint on $\alpha_{\rm CO}$ would get smaller, but still
3 times higher than the ULIRG-like value. Nevertheless, this correction factor is unlikely to be a numerical artifact, as it is theoretically justified from the high velocity dispersions observed in high-redshift gas disks (Burkert et al. 2009).

If we interpreted our sources as merging galaxies rather than clumpy disks, the implied dynamical masses would have grown even higher (see, e.g., Tacconi et al. 2008), and so would the conversion factor.  

We might also consider systematic effects in the determination of
the stellar mass. We have used a bottom-light IMF that is the best choice according to observations 
in the local Universe (Chabrier 2003; see also Kroupa et al. 2002).
Several authors have suggested that more top-heavy IMFs
might be appropriate in the distant Universe (Dav\'e 2008; van Dokkum 2008), which would further decrease the estimated stellar masses and hence increase the conversion factor. 

There remains the possibility that these galaxies could contain large reservoirs of old stars, formed at very high
redshifts, that could be missed because the light is dominated by young stars. In practice, we would need to
postulate hidden stellar mass reservoirs that should increase the total stellar masses by factors of 2--3 in order to
bring our estimate of $\alpha_{\rm CO}$ to the ULIRG-like level. However,
here we
 use observations up to the near-IR rest-frame (from IRAC) that should correctly
retrieve the masses of old stars without such a strong bias. 
Also, these putative old stellar reservoirs would most likely be
in bulges and halos, which actually can {\em not} contain most of the stellar mass around $z\sim 2$ disks
according to the modeling of Bournaud
\& Elmegreen (2009). 
In addition, the existence of the stellar mass-SFR correlations suggests that the appropriate star formation histories for high redshift galaxies have SFRs that are rising rapidly with time (e.g., Renzini 2009) so that the galaxies that we observe formed most of their stars quite recently.

Another assumption that we have made is a
dark matter fraction of 25\% inside the effective radius. There is of course a
plausible range around this value. If we change the estimate of the
 dark matter fraction to 20\% (30\%) the average $\alpha_{\rm
CO}$ would increase (decrease) by about 10\%. 
A survey of local spirals by Pizagno et al. (2005) shows a total-to-stellar mass ratio around two
within 2.2 times the exponential disk scale-length for galaxies with baryonic masses $\simeq 10^{11}$~M$_{\sun}$,
which for an exponential disk corresponds to a total-to-stellar ratio of 1.4 within the half-light radius, i.e. a
dark matter fraction of 20--25\%. Padmanabhan et al. (2004) find similar dark matter fractions in massive galaxies
and emphasize the fact that higher fraction are found only in low-mass and low-density galaxies. The recent mass
models of the Milky Way by Xue et al. (2008) correspond to a dark matter fraction of 23\% within the half-light
radius, using NFW profiles for the halo. There is no known reason for a major change with redshift\footnote{Adiabatic
contraction of halos may actually increase the dark matter mass enclosed by a given radius with time, but the effect should be
weak between $z=1.5$ and $z=0$. }. The observed objects could turn into elliptical galaxies by redshift zero, but the
dark matter fraction still remains limited to less than 25\%, and likely 15\%, within the half-mass radius (e.g.,
Gnedin \& Ostriker 2000). Thus, our choice of 25\% lies on the conservative side of the plausible range. 
A dark matter fraction of 60\%, which would lower our $\alpha_{\rm CO}$ to the ULIRG value, appears to be very unlikely.

\section{Implications on star formation in disk galaxies at high redshift}
\label{sec:discussion}

\subsection{Massive, extended reservoirs of low-excitation molecular gas}

CO[2-1] detections of two near-IR selected (BzK)
galaxies at $z=1.5$ were presented in Daddi et al. (2008). We have now
observed a total of six galaxies in CO[2-1] and detected all of them. This clearly supports
the ubiquity of massive molecular gas reservoirs in these galaxies, a result 
confirmed by Tacconi et al. (2010).

In our observations we were able to spatially resolve robustly the CO[2-1] 
emission in this kind of
galaxies and derive typical CO emission FWHM in the range of 6--11~kpc. This is a factor of 2--3 larger than what is found in SMGs (Tacconi et al. 2006; 2008; Bouch\'e et al. 2008). 
The presence of extended gas disks is quite consistent with our previous finding based on the low CO excitation 
in these objects (Dannerbauer et al. 2009). 
These findings are indicative of colder and/or lower density gas than what is found in SMGs, a situation
more similar to what is found in local spiral galaxies. 

\subsection{Normal star-forming galaxies at high redshift as extended, clumpy rotating disks}

Our sample is based on the BzK-selection technique and we have shown it to be representative of ``normal'', 
typical star-forming galaxies at high redshift. We now summarize the many pieces of evidence that these 
CO-detected BzK galaxies are large disk galaxies rather than on-going violent mergers. 

First, we clearly detect double peaked profiles and the velocity width 
in the blue- and redshifted components is very similar. This would be a
highly contrived coincidence if this was due to a merger: the components would need to have equal mass and 
be observed at similar inclinations.

Second, the UV rest-frame morphology disfavors on-going major mergers: 
in no case do we see evidence for a double-galaxy encounter.  The irregular and clumpy
morphology could lead one to speculate that this is due to the
presence of multiple minor merging events, but in all likelihood we
are looking at clumps in single massive disks 
(Elmegreen et al. 2007, 2009a, Bournaud et al. 2008, Genzel et al. 2008). 

Other evidence for rotating disks
 include the position of these sources in the velocity-size plane. 
Bouch\'e et al. (2007) used H$\alpha$ resolved spectroscopy of optical and near-IR selected galaxies and identified a
dichotomy between SMGs and extended disks in such velocity-size diagrams: SMGs have much higher velocities and smaller
sizes. Using CO velocity widths and CO sizes for both BzK galaxies and SMGs, 
we confirm a dichotomy between the two classes: in the notation of Bouch\'e et al., our CO-detected galaxies have
'maximum rotational velocities' $v_d = v_{FWHM}/2.35 \simlt200$~km~s$^{-1}$ and exponential disk scale lengths
$r_d>3$~kpc, while the SMGs in Bouch\'e et al. have $v_d>200$~km~s$^{-1}$ and $r_d<3$~kpc.
The latter are better understood in terms of merging
systems, while the former appear consistent with being non-interacting disk galaxies (or interacting only with low-mass satellites).

Further evidence that we are looking at disk galaxies comes from
the CO properties. We have already discussed the large spatial sizes
of the CO reservoirs, and the low gas excitation. Furthermore, the SFE
in these galaxies is considerably lower than that of both SMGs and
local ULIRGs, suggesting a different star formation {\em mode}.  The
gas depletion timescales are of order of 0.5~Gyr and imply a high duty-cycle (see also Daddi et al.\ 2005; 2007; 2008) -- this is quite different from the timescales expected for a merger-induced event that would result in a $\simlt100$Myr starburst (Mihos \& Hernquist 1996; Greve et al. 2005; Di~Matteo et al. 2008). 
As can be seen from Fig.~\ref{fig:both_plot_6CO}, SMGs typically also have depletion timescales of $<$100\,Myr.

\subsection{Origin of massive disks at high redshift}

The fact that we are presumably looking at non-interacting rotating disks instead of on-going mergers does not in itself imply
that these massive disks were not assembled by past mergers. Remnants of major mergers can have massive rotating gas
disks, especially when the gas fraction is high (e.g., Robertson \& Bullock 2008). 
Numerical models predict that, under some physical assumptions, merger remnants
can contain extended gas disks (e.g. Hopkins et al. 2009).
%2009MNRAS.397..802H)
However, Bournaud \& Elmegreen
(2009) argued that the overall spatial distribution of the mass in high-redshift disks is 
inconsistent with the majority of them being merger
remnants, suggesting that their mass assembly was dominated by relatively smooth mass infall along gas streams (see
also Dekel et al. 2009b). This is because  the morphology of these disk galaxies, 
in particular their supermassive star-forming clumps, cannot be
accounted for if the main assembly channel is by mergers.
Our CO observations add further support to this picture. Indeed, we find extended CO
reservoirs with spatial sizes consistent with what is measured from their stellar light. The end product of galaxy
mergers are characterized by highly concentrated molecular gas distributions, even when some molecular gas is found
in the outer regions. This holds both for early-stage interactions (pre-merger galaxies like Arp~105 -- Duc et al.
1997), on-going mergers (like NGC~520, Yun \& Hibbard 2001), and merger remnants (like NGC~7252, Hibbard et al.
1994). This dynamical effect results from the gravitational torquing of the gas, which is well reproduced by numerical
simulations (Barnes \& Hernquist 1991; Di~Matteo et al. 2008) and is independent of the gas fraction. If major
mergers played a role in the formation of $z=1.5$ BzK galaxies, their large molecular disks must have been
replenished by the smooth infall of baryons at a rate higher than the merger-driven mass assembly in order to keep
extended gas disks. Cold flow-driven galaxy formation, indeed, does predict a large population of massive, extended
star-forming disks at high redshift with high gas fractions (e.g., Dekel et al. 2009b).

\subsection{Global properties of star formation at high redshift}

In a simple cartoon, the $z=1.5$ BzK galaxies appear to be disks like
local spiral galaxies, but containing a substantially
higher gas content and much larger star-forming clumps. A similar comparison holds for SMGs vs. local ULIRGs, with
the former being the scaled--up version of the latter but having overall higher gas content (Tacconi et al. 2006). 
Most of the observed properties
of the two classes can be understood within this scheme: 
there is a population of disk  
galaxies having low excitation CO, spatially extended gas reservoirs,
low SFE (and hence long gas consumption timescales), low
SSFR and optically thin UV emission. These are the familiar
spirals in the local Universe, and likely most of the near-IR (or
optical) selected disk galaxies in the distant Universe. Conversely, there
are galaxy populations with opposite properties: high CO
excitation, compact gas reservoirs, high SFE, high SSFR and with
optically thick UV emission. These are the local ULIRGs and high redshift
SMGs, both classes likely corresponding to major (wet) mergers of
massive star forming galaxies.

\subsection{New constraints for galaxy formation models and the IMF}

Based on our results of a remarkably low scatter in observed SFE and gas fractions for the CO-detected BzK galaxies,
we suggest that the molecular gas content plays an important role 
in driving the tight stellar mass-SFR correlation in high redshift sources. A simple scenario may be that the gas properties are tightly connected to that of the underlying dark matter halo, possibly via the regulation of cold gas accretion rates through cosmic times (e.g., Keres et al. 2005; see also Bouch\'e et al. 2010). This in turn could regulate the amount of star formation in each halo and its time integral, i.e. the stellar mass, which, as a result, are tightly connected. Ultimately, there might be a stellar mass-SFR correlation because the gas content of galaxies tightly correlates with the masses of hosting dark matter halos over cosmological timescales.

Galaxy formation models based on numerical hydrodynamic codes (e.g., Finlator et al. 2006) and semianalytical
renditions (e.g., Kitzbichler \& White 2007; see Fig.~17~and~18 in Daddi et al. 2007) had indeed predicted the
existence of tight correlations between stellar mass and SFR before they were recovered by observations, 
with roughly a correct slope (or possibly slightly shallower, see, e.g.,
 Daddi et al. 2007a). This is probably one of the rare cases where models of galaxy formation have preceded observations, showing truly predictive power.  Also, such models had qualitatively predicted an increase in the normalization of the correlation toward higher redshifts. 

However, existing models fail in
reproducing the actual magnitude of the normalization increase to
$z\sim2$ (falling short by factors of 3--5 compared to observational
estimates; see e.g. Daddi et al. 2007a; Dav\'e 2008). The SFRs
estimated for distant galaxies are generally too high
compared to the prediction of models, a fact that has been long known
for SMGs (Bough et al.\ 2005).
Dav\'e (2008) suggest that this discrepancy between models and
observations can be solved by adopting substantial evolution in the
stellar initial mass function (IMF). This conclusion was also reached
by Baugh et al.\ (2005) in their modeling of SMGs (i.e., for galaxies with
extreme SFRs of typically 1000~$M_\odot$~yr$^{-1}$ that have space
densities at least
 an order of magnitude lower than that of near-IR selected massive galaxies at
$z\sim2$). However, new theoretical frameworks have recently been
proposed in which feeding of star formation through cold flow
filaments (Dekel et al. 2009a; see also Keres et al. 2005) can more
easily justify high star formation rates in distant galaxies.
In addition, we have also found large gas reservoirs, reaching
$10^{11}M_\odot$
in some cases, i.e. the gas accounts for a major fraction of the
baryons in these sources. We are thus witnessing a phase in galaxy
formation in which massive galaxies were truly gas--dominated systems.  These high gas masses can explain the high
observed SFRs and one does not need to advocate non standard IMFs to reduce the SFR, at least for the normal, near-IR
selected (BzK) galaxies. We note that quantitative predictions
of the models by Oppenheimer \& Dave (2006; 2008) on the gas fractions in galaxies at $z\sim2$ (with similar stellar masses to our CO--detected galaxies) result in expected gas mass fractions of order of 15\%, i.e., much smaller than what we derived. Therefore, our study suggests that the molecular gas fraction in galaxy simulations should be increased in order to reproduce the observed IR luminosities, without necessarily introducing a top heavy IMF.

\section{Summary}
\label{sec:summary}

We have presented results from a CO[2-1] survey of normal, BzK-selected star-forming galaxies at $z \simeq 1.5$. The
analysis of the integrated spectra and the spatially-resolved CO observations was made in combination with
high-resolution optical imaging and multi-wavelength data. 
We also used appropriate numerical simulations of clumpy disk
galaxies (BEE07) to derive dynamical masses and compare them with the observed CO luminosities. The main findings and implications can be summarized as follows:

\begin{itemize}

\item All observed galaxies were detected with high $S/N$ ratios. We detected extended CO reservoirs with typical
sizes of $\sim 6$--11~kpc, and presented evidence for rotation from
spatially-resolved CO observations. This is indicative
of the molecular gas being situated in disks that are similar in size and shape to the UV rest-frame morphology. 

\item We have provided direct evidence for a high $\alpha_{\rm CO}$ conversion factor (relating CO luminosity to
gas mass), based on the dynamical masses of the systems. Our results 
show that the conversion factor of  BzK galaxies is high, similar to that of local spirals.
Our derived $\alpha_{\rm CO}$ is a factor of 4--5 larger than what has been derived for local ULIRGs and typically
used for high redshift SMGs.

\item Our observations imply high gas fractions, $\simeq 50$--65\% of the baryons in our $z \simeq 1.5$ galaxy
sample, and quiescent star formation with relatively low SFE and long gas consumption timescales ($\sim$ 0.5~Gyr).

\item Our results also point to a strong correlation between the molecular gas content and other physical properties
of these sources. Once the stellar mass and SFR of a given galaxy are known, the CO luminosities can be predicted to
within 40\% rms. This and the overall similar SFEs seen in these galaxies and local spirals leads us to propose a
relation between CO luminosity and bolometric luminosity that could allow one to systematically estimate
the CO luminosity of massive disk galaxies, presumably at any redshift $z\simlt2$ as the relation can
fit local and $z=1.5$ observations with an overall scatter of only 0.25~dex.

\item Our sample consists of normal star-forming galaxies at high redshift, and we have shown that it is
representative of the majority of $z>1$ near-IR selected
 star-forming galaxies. Our results shed more light on the nature of these
high-redshift disk galaxies. In particular, the large spatial size of the molecular gas reservoirs suggest that they
did not assemble mostly through past episodes of violent mergers. Instead, our observations are more consistent with
smooth mass infall along cosmic flows 
and subsequent internal evolution as the main drivers of disk galaxy formation and star formation at high-redshift.

\end{itemize}

Based on observations with the IRAM Plateau de Bure
Interferometer. IRAM is supported by INSU/CNRS (France), MPG (Germany)
and IGN (Spain).  
We acknowledge the use of
GILDAS software (http://www.iram.fr/IRAMFR/GILDAS).  We are grateful
to Len Cowie for help in cross-checking the spectroscopic redshift of
BzK-12591 from Keck observations, prior to our CO follow-up.  We
thank Claudia Maraston for providing Chabrier IMF models from her
library of synthetic stellar populations templates. We thank Padeli Papadopoulos for comments 
and the anonymous referee for a constructive report.
This research was
supported by the ERC-StG grant UPGAL 240039.  We acknowledge the funding support of
French ANR under contracts ANR-07-BLAN-0228, ANR-08-JCJC-0008 and ANR-08-BLAN-0274.
Support for this work was provided by NASA, Contract Number 1224666
issued by the JPL, Caltech, under NASA contract 1407. DR acknowledges
support from from NASA through Hubble Fellowship grant HST-HF-51235.01
awarded by the Space Telescope Science Institute, which is operated by
the Association of Universities for Research in Astronomy, Inc., for
NASA, under contract NAS 5-26555.
The work of DS was carried out at Jet Propulsion Laboratory,
California Institute of Technology, under a contract with NASA. 
The simulations were performed using HPC resources from GENCI-CCRT (Grant 2009-042192).

\end{document}